\documentstyle[12pt,epsf]{article}

\def\thefootnote{\fnsymbol{footnote}}

\newcommand{\eq}{\begin{equation}}
\newcommand{\en}{\end{equation}}
\newcommand{\eqa}{\begin{eqnarray}}
\newcommand{\ena}{\end{eqnarray}}

\newcommand{\var}{\varepsilon}

\newcommand{\br}{\langle}
\newcommand{\kt}{\rangle}

\newcommand{\th}[1]{\vartheta_{#1}(\tau)}

\newcommand{\JP}[1]{J.\ Phys.\ {\bf #1}}
\newcommand{\NP}[1]{Nucl.\ Phys.\ {\bf #1}}
\newcommand{\PL}[1]{Phys.\ Lett.\ {\bf #1}}

\newcommand{\PR}[1]{Phys.\ Rev.\ {\bf #1}}

\begin{document}
\begin{titlepage}
\vskip0.5cm
\begin{flushright}
DFTT 58/99\\
HUB-EP-99/42\\
\end{flushright}
\vskip0.5cm
\begin{center}
{\Large\bf Critical amplitudes  and mass spectrum of}
\vskip 0.3cm
{\Large\bf the 2D Ising model in a magnetic field}
\end{center}
\vskip 1.3cm
\centerline{
M. Caselle$^a$\footnote{e--mail: caselle~@to.infn.it}
 and M. Hasenbusch$^b$\footnote{e--mail: hasenbus@physik.hu-berlin.de}}
 \vskip 1.0cm
 \centerline{\sl  $^a$ Dipartimento di Fisica
 Teorica dell'Universit\`a di Torino}
 \centerline{\sl Istituto Nazionale di Fisica Nucleare, Sezione di Torino}
 \centerline{\sl via P.Giuria 1, I-10125 Torino, Italy}
 \vskip .4 cm
 \centerline{\sl $^b$ Humboldt Universit\"at zu Berlin, Institut f\"ur Physik}
 \centerline{\sl Invalidenstr. 110, D-10099 Berlin, Germany}
 \vskip 1.cm

\begin{abstract}
We compute the spectrum and several critical amplitudes of the two
dimensional Ising model in a magnetic field with the transfer matrix 
method. The three lightest masses and their overlaps with the spin and the
energy operators are computed on lattices of a width
up to $L_1=21$. In extracting the continuum results we also take
into account the corrections to scaling due to irrelevant operators. 
In contrast with previous Monte Carlo simulations our final results are
in perfect agreement with the predictions of S-matrix and conformal field 
theory. We also obtain the amplitudes of some of the subleading corrections,
for which no S-matrix prediction has yet been obtained.

\end{abstract}
\end{titlepage}

\setcounter{footnote}{0}
\def\thefootnote{\arabic{footnote}}

\section{Introduction}

In these last years there has been much progress in the study of 
2d spin models in the neighbourhood of critical points. The scaling limit of 
such models is described in general by the action ${\cal A}$
obtained by perturbing 
the conformal field theory (CFT)
which describes the critical point with one (or more) of the 
relevant operators which appear in the spectrum of the CFT.
\eq
{\cal A} = {\cal A}_0 + \lambda \int d^2x \, \phi(x) \,\,\,
\label{action}
\en
where ${\cal A}_0$ is the action of the CFT at the critical point~\cite{bpz}
 and $\phi(x)$
is the perturbing operator. Few years ago A. Zamolodchikov in a seminal
paper~\cite{z89} suggested that in some special cases these perturbed theories
 are equivalent to relatively simple
 quantum field theories~\cite{int}
 whose mass spectrum and S-matrix are explicitly known.
Later it was realized that these theories had a deep connection with the Dynkin
diagrams of suitable Lie algebras and, from the exact knowledge of the S-matrix,
several other informations, and in particular some critical amplitudes were
obtained (for a review, see for instance~\cite{rev}). While these results have
formally the status of conjectures, they successfully passed
 in these last years so many tests that they are now universally accepted. 
The most fascinating example of these S-matrix models is the Ising model
perturbed by an external magnetic field, which is also the model which was
originally studied by Zamolodchikov in~\cite{z89}. This model is highly
non-trivial. Its spectrum contains 8 stable scalar particles, all with different
masses. Both the masses and the entries of the S-matrix are based of the
numerology of the $E_8$ exceptional Lie  algebra. In particular the ratio
between the first two masses is predicted to be the ``golden ratio''
$m_2/m_1=2 \cos(\frac{\pi}{5})$. The simplest realization of this QFT is the 
2d Ising model at $\beta=\beta_c$ in presence of an external magnetic field $h$.
However there are several other models which belong to the same universality
class. In particular, the first numerical check of the predictions of~\cite{z89}
was performed on the Ising quantum spin chain~\cite{hs} in which the 
first few states
of the spectrum were precisely observed. Another interesting realization 
 was presented in~\cite{bnw,bs97},
where  the dilute $A_3$ IRF 
(Interaction Round a Face) model was solved exactly and the predicted spectrum
of states was found~\cite{bs97}. 

Despite these successes, little progress has been achieved in testing 
Zamolodchikov's
proposal directly in the 2d Ising spin model. Even more, it is exactly for 
this model that one faces the only existing 
 discrepancy between Zamolodchikov's results and Monte Carlo simulations.

The Ising spin model in a magnetic field was studied numerically
in~\cite{lr,destri}.
In both papers the authors studied the spin-spin correlator and
did not find the spectrum predicted
by~\cite{z89}. Their data were compatible with the presence in the spectrum of
only the lowest mass state.
 The explanation suggested in~\cite{lr,destri} was that probably the
higher masses had a negligible  overlap amplitude with the spin operator.
However, later, in~\cite{dm} these overlaps were evaluated explicitly in the
S-matrix framework and
turned out to be of the same order of magnitude as the overlap 
with the lowest mass
state.

In this paper we shall address this problem. We shall show that Zamolodchikov's
proposal (and the calculations of~\cite{dm}) is correct also in the case of the
2d Ising spin model and
that the apparent disagreement was due to the fact 
that it is very difficult to extract a complex spectrum from a
multi-exponential fit to the spin-spin correlator.
We have been prompted to this
explanation by another example that we recently studied, in which exactly the
same phenomenon happens: the 3d Ising model~\cite{chp}. In this case also, 
a multi-exponential fit to the spin-spin correlator seems to indicate the
presence of a single state in the spectrum, while using a suitable variational
method and diagonalizing a set of improved operators one can clearly see the
rich spectrum of the model. 

While in previous numerical works~\cite{lr,destri} the model was studied by
using Monte Carlo simulations we tried in the present paper a different
approach based on the exact diagonalization of the transfer matrix.

This approach has various advantages: it gives direct access to the mass
spectrum of the model and allows to obtain numerical estimates of various
quantities with impressively small uncertainties. However it has the
serious drawback  that only transfer matrices 
of limited size can be handled and it is difficult to extract from them 
the continuum limit results in which we are interested. During the last years
various strategies have been elaborated to attack this problem, but all of them
are affected by systematic errors whose size is in general unknown.

In this paper we propose a new approach based on the fact that, by using the
exact solution of the Ising
model at the critical point, one can construct very precise expansions for the
scaling functions in powers of the perturbing field. More precisely, thanks to
the knowledge of the spectrum of the model, it is possible
to list all the irrelevant fields
which may appear in the effective Hamiltonian and select them on the 
basis of the symmetry properties of the observables under study. 

Our strategy  could be summarized as follows.
\begin{itemize}
 \item Choose a set of values of $h$ for which the correlation length is much
 smaller than the maximum lattice size that we can study\footnote{In particular
 we decided to keep the ratio $\frac{\xi}{L_0}<0.1$. This means that we only
 studied values of $h$ for which the correlation length was 
 smaller than two lattice spacings.}. 
 Diagonalize the transfer matrix for various
 values of the transverse size of the lattice and extract all the observables of
 interest.
\item
   Extrapolate the numbers thus obtained to the thermodynamic limit. Thanks to
   the very small correlation length, the finite size behaviour is dominated by 
   a rapidly decreasing exponential and the thermodynamic limit can be reached
   with very small uncertainties (we list in a set of tables at the end of the
   paper the results that we obtained in this way).
\item
   Construct for each observable the scaling function keeping the first 7 or 8
   terms in the expansion in powers of the perturbing field.
\item
   Fit the data with these truncated scaling functions. By varying the 
   the number of input data and of subleading terms used in the scaling
   functions we may then obtain a reliable estimate of the systematic deviations
   involved in our estimates (see the discussion in sect.6). 
\end{itemize}

Which are the observables of interest mentioned above?

Usually, when looking at the scaling regime of statistical models one can study
only adimensional amplitude ratios which are the only quantities which, thanks
to universality, do not depend on the details of the lattice models, but only on
the features of the underlying QFT. 
However the Ising model can be solved
exactly at the critical point also on the lattice and explicit expressions for
 the spin-spin and energy-energy correlators are known. 
This allows to write 
 the explicit expression in lattice units of the amplitudes
 evaluated in the framework of the S-matrix theory. Thus one is able to predict
 not only adimensional amplitude ratios  {\sl 
but also the values of the critical amplitudes themselves}. This greatly enhances
the predictive power of the S-matrix theory and makes much more stringent the
numerical test that we perform.

The final result of our analysis is that all the observables that we can measure
perfectly agree with the S-matrix predictions.

In particular we obtain very precise estimates for the first three
masses, for several critical amplitudes and, what is more important,
 for the overlap
amplitude of the first two masses with the spin and energy operators, a result 
 which had never been obtained before. 

We also measure the amplitude of some of the
subleading corrections in the scaling functions, 
for which no S-matrix prediction exists for
the moment. In particular we found that
the amplitude of the corrections due to the energy momentum tensor in
translationally invariant observables is compatible with zero.

This paper is organized as follows. In sect.~2 we introduce the model in which
we are interested, collect some known results from S-matrix theory and
finally give the translation in lattice units of the critical amplitudes
evaluated in the S-matrix framework. In sect.~3 we construct the scaling
functions.
This only requires the use of very 
simple and well known results of Conformal Field Theory. Notwithstanding
this  it turns out to be a rather non trivial exercise. 
Since it could be a result of general utility (it could be extended
 to other models for which the CFT solution is known or to other 
 quantities of the Ising model in a magnetic field
 that we have not studied in the present paper)
 instead of simply giving the results,
 we derived the scaling functions explicitly and tried to give as much 
details as possible.
 Sect.~4 is devoted to a description of the
transfer matrix method. 
In sect.~5 we deal with 
the thermodynamic limit while in sect.~6 we analyze the transfer matrix
 results and give our
best estimates for the critical amplitudes in which we are interested. 
Sect.~7 is devoted to some concluding remarks.

 To help the reader to reproduce our analysis (or to follow some alternative
fitting procedure) we list in four tables at the end of the paper the data that
we obtained with the transfer matrix approach.

\section{Ising model in a magnetic field}
In this section we shall review the existing theoretical informations on the
Ising model in a magnetic field. This will require four steps. First (in 
sect.~2.1) we shall
define the lattice version of the model, discuss its action and define the 
observables in which we shall be interested in the following. Then 
(in sect.~2.2) we shall 
turn to the continuum version of the theory, described in the present case
by the action:
\eq
{\cal A} = {\cal A}_0 + h \int d^2x \, \sigma(x)  
\label{action2}
\en
where $\sigma(x)$ is the perturbing operator. In particular
we shall  discuss, within the framework of the renormalization group,
the expected scaling behaviour of the 
various quantities of interest and define the corresponding critical
amplitudes. In  sect.~2.3 we
shall use the knowledge of the S-matrix of the model to obtain the value of
some of the amplitudes of interest by using the Thermodynamic Bethe Ansatz (TBA)
 and
the  form factor approach. Finally in sect.~2.4 we shall turn back to the
lattice model and show how the continuum results can be translated in lattice
units.

\subsection{The lattice model}
The lattice version of the Ising model in a magnetic field is defined
by the partition function
\eq
Z=\sum_{\sigma_i=\pm1}e^{\beta(\sum_{\br n,m \kt}\sigma_n\sigma_m
+H\sum_n\sigma_n)}
\label{zz1}
\en
where the field variable $\sigma_n$ takes the values $\{\pm 1\}$;
$n\equiv(n_0,n_1)$ labels the sites of a square lattice of size $L_0$ and $L_1$
in the two directions
and $\br n,m \kt$ 
denotes nearest neighbour sites on the lattice.
In our calculations with the transfer
matrix method we shall treat asymmetrically the two directions. We shall denote
$n_0$ as the ``time'' coordinate and $n_1$ as the space one.
The number of sites  of
the lattice will be denoted by  $N\equiv L_0 L_1$.
 In the thermodynamic limit both $L_0$
and $L_1$ must go to infinity and only in this limit we may recover the results
of the continuum theory. In our actual calculations with the transfer
matrix method we shall study finite values of $L_1$ and then extrapolate
the results to infinity. This extrapolation induces systematic errors which
are the main source of uncertainty of our results, since  the rounding errors
in the transfer matrix diagonalization are essentially negligible. 
In sect.~6 below, we shall discuss these systematic errors 
and estimate their magnitude.

In order to select only the magnetic perturbation, the coupling
 $\beta$ must be fixed to its
critical value
 $$\beta=\beta_c=\frac12\log{(\sqrt{2}+1)}=0.4406868...$$
 by defining  $h_l=\beta_c H$ we end up with 
\eq
Z(h_l)=\sum_{\sigma_i=\pm1}e^{\beta_c\sum_{\br n,m \kt }\sigma_m\sigma_m
+h_l\sum_n\sigma_n}   \;\;\; .
\label{fupart}
\en

 $h_l$ denotes the lattice discretization of the magnetic field $h$ which
appears in the continuum action eq.~(\ref{action2}). It
 must be, for symmetry reasons, an odd function of $h$.

\subsubsection{Lattice operators}
It is useful to define the lattice analogous of the spin and energy
operators of the continuum theory. They will correspond to  linear combinations
of the relevant and irrelevant operators of the continuum theory with suitable
symmetry properties with respect to the $Z_2$ symmetry of the model
(odd for the spin operator and even for the energy one). Near the critical point
this linear combination will be dominated by the relevant operator and the only
remaining freedom will be a conversion constant relating the continuum and 
lattice versions of the two operators (we shall find this constants in 
sect.~2.5).
  The simplest choices for these lattice analogous are
\begin{itemize}
\item Spin operator
\eq
\sigma_l(x)\equiv \sigma_x
\en
i.e. the operator which associates to each site of the lattice the value of the
spin at that site. 
\item Energy operator
\eq
\epsilon_l(x)\equiv \frac14\sigma_x\left(\sum_{y~n.n.~x} \sigma_y\right)
-\epsilon_b  
\en
where the sum runs over  the four nearest neighbour sites $y$  of $x$.
$\epsilon_b$ represents a constant ``bulk'' term which we shall discuss below.
\end{itemize}
The index $l$ indicates that these are the lattice discretizations of the
continuous operators.
We shall denote in the following the normalized sum over all the sites of these
operators simply as
\eq
\sigma_l\equiv\frac1N\sum_x\sigma_l(x) \hskip2cm
\epsilon_l\equiv\frac1N\sum_x\epsilon_l(x)   \;\;\; .
\en

\subsubsection{Observables}

\begin{itemize}
\item {\bf Free Energy}

The free energy is defined as 
\eq
f(h_l)\equiv \frac1N \log(Z(h_l))   \;\;\; .
\en
It is important to stress that $f(h_l)$ is composed by a ``bulk'' term
 $f_b(h_l)$ which is an analytic even function of $h_l$ and
by a ``singular'' part $f_s(h_l)$ 
which contains the relevant informations on the
theory as the critical point is approached. The continuum theory can give 
informations only on $f_s$. 
The value of $f_b(0)$ can be obtained from the exact
solution of the lattice model at $h_l=0, \beta=\beta_c$ (see~\cite{ff})
\eq
f_b=\frac{2G}{\pi}+\frac12 \log{2}=0.9296953982...
\en
where $G$ is the Catalan constant.

\item {\bf Magnetization}

 The magnetization per site $M(h_l)$ is defined as
\eq
M(h_l)\equiv \frac1N \frac{\partial}
{\partial h_l}(\log~Z(h_l)) 
=  \frac1N \br\sum_i \sigma_i \kt \;\;\; .
\en
Hence we have 
\eq
M(h_l)=\br\sigma_l\kt  \;\;\; .
\en

\item {\bf Magnetic Susceptibility}

The magnetic susceptibility $\chi$ is defined as
\eq
\chi(h_l)\equiv \frac{\partial M(h_l)}{\partial h_l}  \;\;\; .
\en

\item {\bf Internal Energy}

We define the internal energy density $\hat E(h_l)$ as
\eq
\hat E(h_l) 
\equiv \frac{1}{2N}\br \sum_{\br n,m \kt}\sigma_n\sigma_m \kt   \;\;\; .
\en
As for the free energy, also in this case one has a bulk analytic contribution
$E_b(h_l)$ which is an even function of $h_l$.
Let us define $\epsilon_b\equiv E_b(0)$.
The value of $E_b(0)$ can be easily evaluated (for instance by
using Kramers-Wannier duality) to be $\epsilon_b=\frac{1}{\sqrt2}$. 
Let us define $E(h_l)\equiv \hat E(h_l)-\epsilon_b$, we have
\eq
E(h_l) = \frac{1}{2N} \br \sum_{\br n,m\kt}\sigma_n\sigma_m
\kt
-\frac{1}{\sqrt{2}}   \;\;\; .
\label{bulk}
\en
Hence we have 
\eq
E(h_l)=\br\epsilon_l\kt   \;\;\; .
\en

As usual the internal energy can also be obtained by deriving the free energy
with respect to $\beta$. However it is important to stress that, due to the
magnetic perturbation (see eq.(\ref{zz1})) in performing the derivative we also
extract from the Boltzmann factor a term proportional to $H\sigma_l$. Hence we
have:
\eq
\hat E(h_l)=
\frac{1}{2N}\frac{\partial}
{\partial \beta}(\log~Z(h_l)) 
-\frac{h_l}{2\beta_c}\sigma_l~~~~~.
\label{new1}
\en
This observation will play an important role in the following.

\end{itemize}

\subsubsection{Correlators}
We are interested in the spin-spin and in the energy-energy connected 
correlators defined as
\eq
G_{\sigma,\sigma}(r)
\equiv \br \sigma_l(0)\sigma_l(r) \kt -\br \sigma_l \kt^2
\equiv \br \sigma_l(0)\sigma_l(r) \kt_c \;\;\; ,
\en
\eq
G_{\epsilon,\epsilon}(r)
\equiv \br \epsilon_l(0)\epsilon_l(r) \kt -\br\epsilon_l\kt^2
\equiv \br \epsilon_l(0)\epsilon_l(r) \kt_c  \;\;\; .
\en
For a nonzero magnetic field
these correlators are very complicated, unknown, functions of $h$ and $r$, 
however a good
approximation in the large distance regime $r\to\infty$ 
is\footnote{For a discussion of the limits of this approximation and of 
the corrections which must be taken into account when the short distance regime
is approached see~\cite{cgm}.}
\eq
\frac{G_{\sigma\sigma}(r)}{\br\sigma_l\kt^2}~~=~~\sum_i
\frac{\left|F^\sigma_i(h)\right|^2}{\pi} K_0(m_i(h) r)
\label{threes}
\en
where  the sum is over the low laying single particle states of the spectrum,
 $m_i(h)$ denotes their mass the functions $F^{\sigma}_i(h)$
 their overlap with the $\sigma$ operator.
Similarly we have
\eq
\frac{G_{\epsilon\epsilon}(r)}{\br\epsilon_l\kt^2}~~=~~\sum_i
\frac{\left|F^\epsilon_i(h)\right|^2}{\pi} K_0(m_i(h) r)
\label{thre2}
\en
where the spectrum is the same as for the spin-spin correlator but the overlap
constants are different. 

A particular role is played by the lowest mass $m_1$ which gives the dominant
contribution in the large distance regime. Its inverse corresponds to the
(exponential) correlation length $\xi$ of the model and sets the scale for all
dimensional quantities in the model. In particular the ``large distance regime''
mentioned few lines above  means ``large with respect to the
correlation length''.

In the approximation of eqs. (\ref{threes}) and (\ref{thre2})
 one is neglecting  the cut-type contributions which appear above the
 two-particle threshold i.e. at twice the value of $m_1$. For this reason we
 shall concentrate in the following only on the three first states of the
 spectrum which are the only ones which lie below  such threshold (see 
eq. (\ref{mass_spectrum}) below).

\subsubsection{Time slice correlators}
It is very useful to study the zero momentum projections of the above defined
correlators. They are commonly named time slice correlators. The magnetization
of a time slice is given by
\eq
S_{n_0}\equiv \frac1{L_1}\sum_{n_1}\sigma_{(n_0,n_1)} \;\;\; .
\en
The time slice correlation function is then defined as
\eq
\label{slicecorr}
G^0_{\sigma\sigma}(\tau)\equiv \sum_{n_0}\{\br S_{n_0} S_{n_0+\tau}\kt-
\br S_{n_0}\kt^2\}
\en
where the index $0$ indicates that this is the zero momentum projection of the
original correlator. Starting from eq.~(\ref{threes})  it
 is easy to show that in the large $\tau$ limit $G^0_{\sigma\sigma}(\tau)$
 behaves as
\eq
\frac{G^0_{\sigma\sigma}(\tau)}{\br\sigma_l\kt^2}~~=~~\sum_i
\frac{\left|F^\sigma_i(h)\right|^2}{m_i(h) L_1} e^{-m_i(h) |\tau|} \;\;\; .
\label{threes0}
\en
A similar result, with the obvious modifications,
 holds also for $G^0_{\epsilon,\epsilon}$.

\subsection{Critical behaviour}
In this section we  discuss the critical behaviour of the model 
 by using standard renormalization group methods,
keeping in the expansions
only  the first order in the perturbing field. Both the results and the
 analysis are well known and can be found in any textbook. 
We report it here since it will serve us as a  starting point
 for the more refined analysis which we shall perform in sect.~3 below.

\subsubsection{Critical indices}
The starting point of the renormalization group
 analysis is the singular part of the free energy
$f_s(t,h)$ (where $t$ is the reduced temperature).
 Standard renormalization group arguments (see for instance~\cite{cardy})
 allow to write $f_s$ in
terms of a suitable scaling function $\Phi$:
\eq
\label{a1}
f_s(t,h)=\left\vert\frac{u_h}{u_{h_0}}\right\vert^{d/y_h}
 \Phi\left(\frac{u_t/u_{t_0}}{|u_h/u_{h_0}|^{y_t/y_h}}\right)
\en
 where  $u_{t_0}$ and
 $u_{h_0}$ are reference scales that depend on the model.
$u_h$ and $u_t$
 denote the scaling variables associated to 
 the magnetic and energy operators respectively
 and
 $y_h$, $y_t$ are their RG-exponents.
 $u_t$ and
$u_h$ do not exactly coincide with $t$ and $h$ but are instead  
analytic functions of them. The only 
 constraint is that they must
 respect the $Z_2$ parity of $t$ and $h$ .
Near the critical point we may suitably rescale $\Phi$ so as to identify
$u_t=t$ and $u_h=h$. Thus, 
setting $t=0$ we immediately obtain the 
 asymptotic critical behaviour of $f_s$
\eq
f_s\propto |h|^{d/y_h}~~~.
\label{asy1}
\en

Taking the derivative with respect to $h$ (or $t$) and then 
setting $t=0$ we can obtain from eq.~(\ref{a1}) also the 
 asymptotic critical behaviour of the other observables
 in which we are interested 
\eq
M\propto |h|^{d/y_h-1}
\label{asy2}
\en
\eq
\chi\propto |h|^{d/y_h-2}
\label{asy3}
\en
\eq
E\propto |h|^{(d-y_t)/y_h}  \;\;\; .
\label{asy4}
\en

From the exact solution of the Ising model at the critical point we know that
$y_h=\frac{15}{8}$ 
and $y_t=1$. Inserting these values in the above expressions we
find

\eq
f_s\propto |h|^{\frac{16}{15}}
\label{asy1b}
\en
\eq
M\propto |h|^{\frac{1}{15}}
\label{asy2b}
\en
\eq
\chi\propto |h|^{-\frac{14}{15}}
\label{asy3b}
\en
\eq
E\propto |h|^{\frac{8}{15}}  \;\;\; .
\label{asy4b}
\en

The masses $m_i$ have as scaling exponent, as usual, $1/y_h$, hence
\eq
m_i\propto |h|^{\frac{8}{15}}  \;\;\; .
\label{asy5b}
\en

Finally from the definitions of eqs.~(\ref{threes}) and (\ref{thre2})
 we see that the
overlap amplitudes behave as adimensional constants.

\subsubsection{Critical amplitudes}
In order to describe the scaling behaviour of the
model we also need to know the proportionality constants in the above
scaling functions.
These constants   are usually called critical amplitudes.
Using the results collected in eqs.~(\ref{asy1b})-(\ref{asy4b}) we
 have the following definitions:

\eq
A_f \equiv \lim_{h\to 0} f~h^{-\frac{16}{15}}~,~~~~~~~
A_M \equiv \lim_{h\to 0} M~h^{-\frac{1}{15}}~,~~~~~~~
A_\chi \equiv \lim_{h\to 0} \chi~h^{\frac{14}{15}}~,
\label{defcrit1}
\en
\eq
A_E \equiv \lim_{h\to 0} E~h^{-\frac{8}{15}}~,~~~~~~~
A_{m_i} \equiv \lim_{h\to 0} m_i~h^{-\frac{8}{15}}~,
\en
\eq
A_{F_i^\sigma} \equiv \lim_{h\to 0} F^{\sigma}_i~,~~~~~~~~
A_{F_i^\epsilon} \equiv \lim_{h\to 0} F^{\epsilon}_i~~.
\en

Notice for completeness that in the literature (see for
 instance~\cite{ahp}) the
 amplitudes $A_\chi$, $A_M$ and $A_{m_1}$ are usually denoted as
\eq
A_\chi\equiv \Gamma_c~,~~~
A_M\equiv D_c^{-\frac{1}{15}}~,~~~
A_{m_1}\equiv\frac{1}{\xi_c}  \;\;\; .
\label{not1}
\en

We shall show in the next section that all these amplitudes can be exactly
evaluated in the framework of the $S$-matrix approach. 
As a preliminary step let us notice that
since $M$ and $\chi$ are obtained as derivatives of $f$
 we have
\eq
A_M=\frac{16}{15} A_f~,~~~~~~~~
A_\chi=\frac{1}{15} A_M~ \;\;\; .
\en

\subsubsection{Universal amplitude ratios}
From the above critical amplitudes one can construct universal
 combinations which do not depend on the particular realization of the 
model. For this reason they have been widely studied in the literature. In
particular there are two ``classical'' amplitude combinations which involve the 
critical amplitudes defined above (see for instance~\cite{ahp}). They are:
\eq
R_\chi\equiv\Gamma D_c B^{14}~~~~~~~~~~
Q_2\equiv (\Gamma/\Gamma_c)(\xi_c/\xi_0)^{\frac{7}{4}}
\en
where we used the notations of 
eq.~(\ref{not1}). $\Gamma$ and $\xi_0$ denote the critical amplitudes of the
susceptibility and exponential correlation length for $h=0$ and a small
 positive reduced temperature, while $B$ denotes the critical amplitude of the
 magnetization for $h=0$ and a small negative reduced temperature.
Notice however that, since (as we mentioned in the introduction) 
we are able to give the
explicit relation between lattice and continuum expectation values, we
 are not constrained to study  only universal combination but can determine
exactly the various critical amplitudes.

\subsection{S-matrix results}

 In 1989 A. Zamolodchikov~\cite{z89} suggested  that
 the scaling limit of the Ising Model in a
magnetic field  could be described by a
a scattering theory which 
contains eight different species of self-conjugated particles
$A_{a}$, $a=1,\ldots,8$ with masses
\eqa
\label{mass_spectrum}
m_2 &=& 2 m_1 \cos\frac{\pi}{5} = (1.6180339887..) \,m_1\,,\nonumber\\
m_3 &=& 2 m_1 \cos\frac{\pi}{30} = (1.9890437907..) \,m_1\,,\nonumber\\
m_4 &=& 2 m_2 \cos\frac{7\pi}{30} = (2.4048671724..) \,m_1\,,\nonumber \\
m_5 &=& 2 m_2 \cos\frac{2\pi}{15} = (2.9562952015..) \,m_1\,,\\
m_6 &=& 2 m_2 \cos\frac{\pi}{30} = (3.2183404585..) \,m_1\,,\nonumber\\
m_7 &=& 4 m_2 \cos\frac{\pi}{5}\cos\frac{7\pi}{30} = (3.8911568233..) \,m_1\,,
\nonumber\\
m_8 &=& 4 m_2 \cos\frac{\pi}{5}\cos\frac{2\pi}{15} = (4.7833861168..) \,m_1\, 
\nonumber
\ena
where $m_1(h)$ is the lowest mass of the theory. 
As mentioned above it coincides with
the inverse of the (exponential) correlation length. Few years later,
from the knowledge of the S-matrix of the theory V. Fateev~\cite{Fateev}
obtained explicit predictions for some of the critical amplitudes defined above.

In order to evaluate the amplitudes 
one must first fix the normalization of the operators involved 
 which can be set, for instance, by fixing the constant in front
of the long distance behaviour of
 the correlators at the critical point. 
It is important to make explicit this normalization choice,
 since it will allow us, by
comparing with  the corresponding correlators in the lattice theory  to convert
 explicitly the continuum results in lattice units.
Following the choice of~\cite{Fateev} we
 assume:
\eq
\br \sigma(x)\sigma(0) \kt =\frac{1}{\,\,|x|^{\frac{1}{4}}}\,\,,
\hspace{1cm}|x|\rightarrow
\infty
\label{uv}
\en
\eq
\br \epsilon(x)\epsilon(0)\kt=\frac{1}{\,\,|x|^{2}}\,\,,
\hspace{1cm}|x|\rightarrow
\infty.
\label{uve}
\en

With these conventions one finds~\cite{Fateev}:

 \eq
A_{m_1} \,=\, {\cal C} 
\label{fat}
\en
\eq
A_f=\frac{{\cal C}^2}{8\,(\sin\frac{2\pi}{3}+
\sin\frac{2\pi}{5}+\sin\frac{\pi}{15})}\, 
\label{sigmah}
\en
where
\eq
{\cal C} \,=\,
 \frac{4 \sin\frac{\pi}{5} \Gamma\left(\frac{1}{5}\right)}
{\Gamma\left(\frac{2}{3}\right) \Gamma\left(\frac{8}{15}\right)}
\left(\frac{4 \pi^2 \Gamma\left(\frac{3}{4}\right)
\Gamma^2\left(\frac{13}{16}\right)}{\Gamma\left(\frac{1}{4}\right)
\Gamma^2\left(\frac{3}{16}\right)}\right)^{\frac{4}{5}} \,
 \,=\, 4.40490858...  \;\; .
\en
From $A_f$ one immediately obtains $A_M$ and $A_\chi$.

The amplitude $A_E$ requires a more complicated analysis.
Its exact expression
 has been obtained only recently in \cite{flzz}
\eq
A_E=2.00314... \;\; .
\en

We summarize in tab.~\ref{tab1} these S-matrix predictions for the critical
amplitudes.

\begin{table}[h]
\begin{tabular}{|ll|}\hline
$ A_{m_1}$ & $=\,\,\,\, 4.40490858.. $ \\
$ A_{f} $ & $=\,\,\,\,   1.19773338.. $ \\
$ A_{M}$ & $=\,\,\,\,   1.27758227..$ \\
$ A_{\chi}$ &$ =\,\,\,\,0.08517215.. $ \\
$ A_{E}$ & $=\,\,\,\, 2.00314.. $ \\
 \hline
\end{tabular}
\vskip 0.2cm
\caption{\sl Critical amplitudes. }
\label{tab1}
\end{table}

From these critical amplitudes, and using the values of $\Gamma,B$ and $\xi_0$
 one immediately obtains the classical 
amplitude ratios defined above (see for instance~\cite{d98}). 
They are reported in tab.~\ref{tab2}~.

\begin{table}[h]
\vskip 0.2cm
\begin{tabular}{|c|}\hline
$ R_\chi = 6.77828502.. $  \\
$ Q_2 = 3.23513834.. $  \\ \hline
\end{tabular}
\vskip 0.2cm
\caption{\sl Classical amplitude ratios. }
\label{tab2}
\end{table}

Finally, the critical overlap amplitudes
$A_{F_i^\sigma}$ and $A_{F^\epsilon_i}$
were  evaluated in \cite{dm,ds}.
They are reported in tab.~\ref{tab3} and~\ref{tab4}~.

\begin{table}[h]
\vskip 0.2cm
\begin{tabular}{|c|}\hline
$ A_{F^{\sigma}_1} =-0.64090211.. $ \\
$ A_{F^{\sigma}_2} =\,\,\,\, 0.33867436.. $ \\
$ A_{F^{\sigma}_3} =-0.18662854.. $ \\
$ A_{F^{\sigma}_4} =\,\,\,\, 0.14277176.. $ \\
$ A_{F^{\sigma}_5} =\,\,\,\, 0.06032607.. $ \\
$ A_{F^{\sigma}_6} =-0.04338937.. $ \\
$ A_{F^{\sigma}_7} =\,\,\,\, 0.01642569.. $ \\
$ A_{F^{\sigma}_8} =-0.00303607.. $ \\ \hline
\end{tabular}
\vskip 0.2cm
\caption{\sl Critical overlap amplitudes for the spin operator. }
\label{tab3}
\end{table}

\begin{table}[h]
\vskip 0.2cm
\begin{tabular}{|c|}\hline
$ A_{F^{\var}_1} =-3.70658437.. $ \\
$ A_{F^{\var}_2} =\,\,\,\, 3.42228876.. $ \\
$ A_{F^{\var}_3} =-2.38433446.. $ \\
$ A_{F^{\var}_4} =\,\,\,\, 2.26840624.. $ \\
$ A_{F^{\var}_5} =\,\,\,\, 1.21338371.. $ \\
$ A_{F^{\var}_6} =-0.96176431.. $ \\
$ A_{F^{\var}_7} =\,\,\,\, 0.45230320.. $ \\
$ A_{F^{\var}_8} =-0.10584899.. $ \\ \hline
\end{tabular}
\vskip 0.2cm
\caption{\sl Critical overlap amplitudes for the energy operator. }
\label{tab4}
\end{table}

\subsection{Conversion to lattice units}

While the values listed in tab.~\ref{tab2},~\ref{tab3} and~\ref{tab4} 
 are universal, 
 the amplitudes listed in tab.~\ref{tab1} depend on the
details of the  regularization scheme. Thus some further work is needed to
obtain their value on the lattice. We shall denote in the following the lattice
critical amplitudes with an index $l$.
Thus, for instance,
\eq
A^l_M=\lim_{h_l\to 0} \br \sigma_l\kt ~h_l^{-\frac{1}{15}}~,~~~~~~~
\en
to be compared with the continuum critical amplitude defined in eq.
(\ref{defcrit1})
\eq
A_M=\lim_{h\to 0} \br \sigma \kt~ h^{-\frac{1}{15}}~.~~~~~~~
\en

In order to relate the lattice results with the continuum ones we must study
the relationship between the lattice operators and the continuum ones. In
general the lattice operators will be given by the most general combination of
continuum operators compatible with the symmetries of the lattice operator
multiplied by the most general analytic functions of $t$ and 
$h$ (with a parity which
is again constrained by the symmetry of the operators involved).
Thus, for instance, anticipating the discussion that we shall make in 
sect.~3, we have
\eq
\sigma_l=f^\sigma_0(t,h)\sigma + f_i(t,h) \phi_i 
\en
where $f^\sigma_0(t,h)$ and $f_i(t,h)$ are suitable
 functions of $t$ and $h$ and with 
$\phi_i$ we denote all the other fields of the theory (both relevant and
irrelevant) 
which respect the  symmetries of the lattice.

A similar relation also holds for the energy operator:
\eq
\epsilon_l=g^\epsilon_0(t,h)\epsilon + g_i(t,h) \phi_i ~~~~~.
\en

 Finally, also $h_l$ is
related to the continuum magnetic field $h$ by a relation of the type
\eq
h_l=b_0(t,h) h
\label{xconvh}
\en
where $b_0(t,h)$ must be an even function of $h$ .
  
At the first order in $t$ and $h$ these combinations greatly simplify and  
essentially reduce to a different choice of normalization between the continuum
operators and their lattice analogous:
\eq
 \sigma_{l} \ \equiv \ R_{\sigma}  \sigma~,~~~~~~~
 \epsilon_{l} \ \equiv \ R_{\epsilon}  \epsilon~,~~~~~
 h_{l} \ \equiv \ R_{h}  h~
\en
where $R_{\sigma}$, $R_{\epsilon}$ and $R_{h}$ are three 
constants which correspond to the
$h\to 0$, $t\to 0$ limit of the $f^\sigma_0$, $g^\epsilon_0$ 
and $b_0$ functions.
 
If we want to compare the S-matrix results discussed in the previous section 
with our lattice results we must  fix these
 normalizations\footnote{This essentially
 amounts to measure all the quantities in units of the lattice spacing. For this
 reason we can fix in the following the lattice spacing to 1 and neglect it.}.
 The simplest way to do
this is to look at the analogous of eqs.~(\ref{uv},\ref{uve}) at 
 the critical point (namely for $h_l=0$)~\cite{smilga} . 

In fact, if $h_l=0$ it is possible to obtain an explicit expression 
for the spin-spin and energy-energy correlators (for a comprehensive review see
for instance~\cite{mccoy}) directly on the lattice, for any value of $\beta$.
Choosing in particular $\beta=\beta_c$, and looking at the large
distance behaviour of these lattice correlators we may immediately fix the
normalization constants. Let us look first at $R_\sigma$.

We know from ~\cite{wu} that:
\eq
 \label{def2}
\br \sigma_i \sigma_j \kt_{h=0}\ = \ \frac{ R_{\sigma}^2}
{|r_{ij}|^{1/4}}
\en
where  $r_{ij}$ denotes the distance on the lattice between the sites $i$ and
$j$ and
\eq
R^2_{\sigma}=e^{3\xi'(-1)}2^{5/24}=0.70338... \;\;.
\en

By comparing this result with eq.~(\ref{uv}) we find
\eq
 R_{\sigma}  = 0.83868... \;\;.
\label{rss}
\en

From this we can also obtain the normalization of the lattice 
magnetic field which must exactly compensate that of the spin operator in the 
perturbation term 
$h\sigma$.
We find:

\eq
R_h \ = \ (R_{\sigma})^{-1}  = 1.1923... \;\;.
\label{ssh}
\en

Combining these two results we obtain the value in lattice units of the constant
$A_\sigma$
\eq
A_M^l=(R_\sigma)^{16/15}A_M=1.058... \;\;.
\label{magl}
\en
From this one can easily obtain also $A^l_f$, $A^l_\chi$ and $A_{m_1}$.

Let us look now at $R_{\epsilon}$.
In the case of the energy operator the connected correlator on the lattice, 
at $h_l=0$ and for any value of $\beta$ has the following
expression~\cite{h67}:
\eq
\br \epsilon_l(0)\epsilon_l(r) \kt_c=\left(\frac{\delta}{\pi}\right)^2
\left[ K_1^2(\delta r)- K_0^2(\delta r)
\right]
\en
where $K_0$ and $K_1$ are modified Bessel functions, $\delta$ is a parameter
related to the reduced temperature, defined as
\eq
\delta=4|\beta-\beta_c|
\en
and with the index $c$ we denote the connected correlator (notice that thanks to
the definition (\ref{bulk}) no disconnected part must be  subtracted at the
critical point and the index $c$ becomes redundant). This expression has a
finite value in the $\delta\to0$ limit (namely at the critical point). In fact
the Bessel functions difference can be expanded in the small argument limit as
\eq
\left[ K_1^2(\delta r)- K_0^2(\delta r)
\right]=\frac{1}{(\delta r)^2}+...
\en
thus giving, exactly at the critical point:
\eq
\br \epsilon_l(0)\epsilon_l(r)\kt =\frac{1}{(\pi r)^2}~~~~~.
\label{defe}
\en
By comparing this result with eq.~(\ref{uve}) we find
\eq
 \ R_{\epsilon}  =  \frac{1}{\pi}
\label{ss2}
\en
and from this we obtain the expression in lattice units of $A_\epsilon$
\eq
A_\epsilon^l=(R_\sigma)^{8/15}(R_\epsilon)A_\epsilon=0.58051... \;\;\;.
\en

Our results are summarized in tab.~\ref{tab5} .

\begin{table}[h]
\vskip 0.2cm
\begin{tabular}{|ll|}\hline
$ A^l_{m_1} $ &$ =\,\,\,\,  4.01039911...
 $ \\
$ A^l_{f} $ &$ =\,\,\,\,        0.99279949...     
$ \\
$ A^l_{M} $ &$ =\,\,\,\,        1.05898612...     
$ \\
$ A^l_{\chi} $ &$=\,\,\,\,     0.07059907...
$ \\
$ A^l_{E} $ &$ =\,\,\,\,        0.58051... $ \\
 \hline
\end{tabular}
\vskip 0.2cm
\caption{\sl Critical amplitudes in lattice units. }
\label{tab5}
\end{table}

\subsubsection{Alternative derivation of $R_\epsilon$}
In this section we discuss, for completeness, an alternative derivation of
$R_\epsilon$. It can be used in those cases in which the correlators are
not known, but the internal energy is known on a finite size lattice at the
critical point. Then $R_\epsilon$ can be obtained by comparing the finite size
behaviour of the internal energy on the lattice with that predicted by
 conformal field theory in the continuum. In the case of the Ising model,
 thanks to the beautiful work by Ferdinand and Fisher~\cite{ff},
 we know that on a square
 lattice of size $L_0 \times L_1$ with $L_0>L_1$ with periodic boundary conditions
 the internal energy must scale as:
\eq
\br\epsilon_l\kt=\frac{\th{2}\th{3}\th{4}}{\th{2}+\th{3}+\th{4}}\frac{1}{L_1}
\label{re1}
\en
where  $\th{i}$ denotes the $i^{\rm th}$ Jacobi theta function and $\tau\equiv
i\frac{L_0}{L_1}$.

The same behaviour can be studied in the continuum theory, by using CFT
techniques. The result~\cite{dsz} is
\eq
\br\epsilon\kt=\frac{\th{1}'}{\th{2}+\th{3}+\th{4}}\frac{1}{L_1}~~~.
\label{re2}
\en
By using the relation
\eq
{\th{1}'}=~\pi~\th{2}\th{3}\th{4}
\en
which allows to express the derivative of  the $\th{1}$ 
in terms of ordinary theta functions we see that the two equations
(\ref{re1}) and (\ref{re2}) agree only if we choose, as we did in the previous
section, $R_\epsilon=\frac1\pi$.

\section{Scaling functions}
In this section we shall construct the scaling functions for the various
quantities in which we are interested. Our aim is  to 
 give the form (i.e. the
value of the scaling exponents) of the
first 7-8 terms of the expansion in powers of $h$ of the scaling functions and
at the same time to identify the operators in the lattice
Hamiltonian from which they originate. To this end we shall first
deal in sect.~3.1
 with the theory at the critical point. 
We shall in particular discuss its spectrum, which can be constructed
explicitly by using CFT techniques. Next, in sect.~3.2, we discuss in the
framework of the renormalization group approach the origin of the
 subleading terms in the scaling functions, and show how to obtain their
 exponents
from the knowledge
of the renormalization group eigenvalues  $y_i$ of the irrelevant operators.
While in general this analysis is only of limited interest since 
the $y_i$ of the irrelevant operators are unknown,
in the present case, thanks to the CFT solution discussed in sect.~3.1,
it becomes highly predictive and will allow us to explicitly
 construct in sects.~3.3 and 3.4 the scaling functions. In particular in 
sect.~ 3.3 we shall 
list all the irrelevant operators which may appear in the effective
 Hamiltonian and discuss their symmetry properties, while in sect.~3.4 we shall
write  the scaling functions and identify the operators involved in the various
 scaling terms.

\subsection{The Ising model at the critical point}

The Ising model at the critical point is described 
 by the unitary minimal model with central charge
$c=1/2$~\cite{bpz}.
Its spectrum  can be divided into three conformal families characterized by
different transformation 
properties under the dual and $Z_2$ symmetries of the model. They
 are the identity, spin and energy families and are
 commonly denoted as  $[{I}],~[\sigma],~[\epsilon]$.
Let us discuss their features in detail.
\begin{itemize}
\item{\bf Primary fields}

  Each family contains a relevant
operator which is called primary field (and gives the name to the entire
family).  
 Their conformal weights are $h_{I}=0$,
 $h_\sigma=1/16$ and  $h_\epsilon=1/2$ respectively. 
The relationship between conformal
 weights and renormalization  group eigenvalues is: $y=2-2h$.
 Hence the
 relevant operators must have $h<1$.

\item{\bf Secondary fields}

All the remaining operators of the three
 families (which are called secondary fields) are generated from the primary
 ones by applying the generators $L_{-i}$ and $\bar L_{-i}$ 
of the Virasoro algebra. In the following
we shall denote the most general irrelevant field in the $[\sigma]$ family
(which are odd with respect to the $Z_2$ symmetry) with the notation
$\sigma_i$ and  the most
general fields belonging to the energy $[\epsilon]$ or to the identity $[I]$
families (which are $Z_2$ even) with $\epsilon_i$ 
and $\eta_i$ respectively.
 It can be shown that by applying  
 a generator of index $k$: $L_{-k}$ or $\bar L_{-k}$  to a field $\phi$ 
 (where $\phi=,{I},\epsilon,\sigma$ depending on the case)
 of conformal weight $h_\phi$  we obtain a new operator of weight
 $h=h_\phi+k$. In general any combination of $L_{-i}$ and $\bar L_{-i}$
 generators is allowed, and the conformal weight of the resulting operator will
 be shifted by the sum of the indices of the generators used to create it.
 If we denote by $n$ the sum of the indices of the generators of type $L_{-i}$
 and with $\bar n$ the sum of those of type $\bar L_{-i}$ the conformal 
 weight of the resulting operator will be $h_\phi+n+\bar n$.
 The corresponding RG eigenvalue will be $y=2-2h_\phi-n-\bar n$, hence all the
 secondary fields are irrelevant operators.

\item{\bf Nonzero spin operators}

 The secondary fields may have a non zero
spin, which is given by the difference  $n-\bar n$. In general one is
interested in scalar quantities and hence in the subset of those irrelevant
fields which have $n=\bar n$. However on a square lattice the rotational 
group is broken to the finite subgroup $C_4$ (cyclic group of order four).
Accordingly, only spin $0,1,2,3$ are allowed on the lattice.
If an operator $\phi$ of the continuum theory has  spin $j\in {\bf N}$,
then its lattice discretization $\phi_l$ behaves as a 
spin $j~({\rm mod}~4)$
operator with respect to the $C_4$ subgroup. As a consequence all the operators
which in the continuum limit have spin $j=4N$ with $N$ non-negative 
integer can appear in the lattice discretization of a scalar field. This will
play a major role in the following.

\item{\bf Null vectors}

 Some of the secondary fields disappear from the spectrum due to the null vector
 conditions. This happens in particular for one of the two states at level 2 in
 the $\sigma$ and $\epsilon$ families and for the unique state at level 1 in
 the identity family. From each null state one can generate, by applying the
 Virasoro operators a whole family of null states hence at level 2 in the
 identity family there is only one surviving secondary field, which can be
 identified with the stress energy tensor.

\item{\bf Secondary fields generated by $L_{-1}$}

 Among all the secondary fields a particular role is played by those generated
 by the $L_{-1}$ Virasoro generator. $L_{-1}$ is the generator of 
 translations on the lattice and as a consequence it has zero eigenvalue on
 translational invariant observables. Another way to state this results is to
 notice that $L_{-1}$ can be represented
 as a total derivative, and as such it gives zero if applied to an operator
 which can be obtained as the integral over the whole lattice of a suitable
 density (i.e. a translationally invariant operator).

\end{itemize}

\subsection{RG analysis for $h\ne 0$}
 We shall discuss the higher order corrections to the RG analysis of sect.~2.2
along the lines of~\cite{af}, to which we refer for a more detailed 
discussion. The only improvement that we make with respect
to~\cite{af} is in the part devoted to the contribution due to the
 irrelevant operators,
in which we shall make use of the results discussed in the previous section.

We expect three types of corrections to the asymptotic results reported in
sect.~2.2:
\begin{description}
\item{a] }
{\bf Analytic corrections.}

 They are due to the fact, already mentioned in sect.~2.2,
 that the actual scaling
variables in the RG approach are not $h_l$ and $t$ but $u_h$ and $u_t$
 which are in principle the most general analytic
 functions of $h_l$ and $t$ which 
respect the $Z_2$ parity of $h_l$ and $t$. Let us write the Taylor expansion
for $u_h$ and $u_t$, keeping only those first few orders
that are needed for our analysis (we use the notations of~\cite{af}).

\eq
u_h~=~h_l~[1~+~c_ht~+~d_ht^2~+~e_h h_l^2~+~O(t^3,th_l^2)]
\label{u_h}
\en
\eq
u_t~=~ t ~+~b_t h_l^2 ~+~c_tt^2 ~+~d_t t^3 ~+~e_tth^2_l~+~ f_t h_l^4 ~+~
 O(t^4,t^2h_l^2)
\label{u_t}
\en
The corrections induced by the higher terms in 
$u_h$ and $u_t$ are of three types.
\begin{itemize}
\item
The first one is very simple to understand.
It is due to the higher powers of $h_l$ contained in $u_h$ which lead to
 corrections to the  power behaviours listed in sect.~2.2.1
which are shifted by even integer powers of $h_l$. For instance in the 
free energy, as a consequence of the $e_h h_l^2$ term in $u_h$, we 
expect a correction of this type:
\eq
f_s(h_l)= A_f^l |h_l|^{\frac{16}{15}}(1+ A_{f,3}^l |h_l|^2+....) \;\;\; .
\en
with $A_{f,3}^l= \frac{16}{15} e_h$. The indices $f,3$ in $A_{f,3}$ only
denote the fact (that we shall discuss in detail in the next section)
 that this term is the third term in the $h_l$ expansion of the 
 scaling function of the singular part of the free energy.

\item
The second type of correction is due to the terms that depend on $h_l$ which
appear in $u_t$. Their peculiar feature is that,
 even if they are originated by analytic terms in the scaling variables, they
 lead in general to non analytic contributions in the scaling functions
For instance, as a consequence of the $b_t h_l^2$ term in $u_t$, we find in the
free energy a correction of the type:

\eq
f_s(h_l)= A_f^l |h_l|^{\frac{16}{15}}
(1+ A_{f,2}^l |h_l|^{2-\frac{y_t}{y_h}}+....) \;\;\; .
\en
 with  $A_{f,2}=\frac{\Phi'(0)}{\Phi(0)} b_t$ and
${2-\frac{y_t}{y_h}}=\frac{22}{15}$.

\item
The  corrections of the third type
 only appear when studying the internal energy.
They are due to the terms linear in $t$ which are present in $u_h$ and 
$u_t$. The most important of these contributions is the one
 due to the $c_ht$ term in $u_h$ which
gives a correction proportional to $h_l^\frac{8}{15}$ to the dominant scaling
behaviour of the internal energy. We shall discuss these terms in sect.~3.4.3
below.
\end{itemize}

\item{b] }
{\bf Corrections due to irrelevant operators in the lattice Hamiltonian.}

 These can be treated within the
framework of the RG as follows. Let us study as an example the case of an
irrelevant operator belonging to the Identity family.
Let us call the corresponding scaling variable
 $u_3$ and its RG eigenvalue $y_3$ (since $u_3$ is irrelevant,
$y_3<0$). In this case the dependence of $u_3$ on $t$  and $h_l$ is 
\footnote{In general for the irrelevant operators
 there is no need  to tune $u_3^0$ to 
$0$ to approach the critical point. 
However we shall see below that, for symmetry reasons, $u_3^0=0$ for all
the irrelevant operators belonging to the $[\sigma]$ and $[\epsilon]$ 
families.}
\eq
\label{a3}
u_3=u_3^0+a t+ b h_l^2 +\cdots  \;\;\; .
\en
Let us for the moment neglect higher
order terms and assume $u_3=u_3^0$. Then looking again at the singular part of
the free energy we find
\eq
\label{a2}
f_s(t,h_l)=|h_l/h_0|^{d/y_h} \Phi(\frac{t}{|h_l|^{y_t/y_h}},
u_3^0|h_l|^{|y_3|/y_h})
\;\;\; .
\en
Since $u_3^0|h_l|^{|y_3|/y_h}$ is small as $h_l \to 0$ 
it is reasonable to assume
that we can expand $f_s$ in a Taylor series of $u_3^0|h_l|^{|y_3|/y_h}$ (notice
that in eq.~(\ref{a2}) $f_s$ is not singular since it is evaluated at 
$|h_l|>0$). Hence we find (setting again $t=0$)
\eq
f_s= |h_l|^{d/y_h}(a_1+a_2u_3^0|h_l|^{|y_3|/y_h}+\cdots)
\label{uf1}
\en
where $a_1$, $a_2$, $u_3^0$ are non-universal constants.

 This  analysis  can be repeated without changes
for any new irrelevant operator: $u_4$, $y_4$ and so on.
As a last remark, notice  that
on top of these non analytic corrections  we also expect analytic
contributions due to the higher order terms contained in eq.~(\ref{a3}).

While in general this analysis is only of limited interest since 
the $y_i$ of the irrelevant operators are usually unknown,
in the present case
we may identify the irrelevant operators with the secondary fields
 discussed in 3.1  and  use the corresponding RG-exponents as 
input of our analysis.

\item{c] }
{\bf Logarithmic corrections.}

As it is well known, the specific heat of the 2d Ising model at $h_l=0$ 
approaches the critical point with a logarithmic singularity. This means that
in the free energy there must be a term of the type $\Phi_0 u_t^2 log(u_t)$.
While in general we could expect $\Phi_0$ to be a generic function of the ratio
 $u_t/u_h^{8/15}$, the absence of leading log corrections in $M$ and
$\chi$ strongly constraints this function which is usually assumed to be a
simple constant. Notwithstanding this, the presence of terms that depend on 
$h_l$ in $u_t$ implies that log type contributions may appear also in the  case
$t=0$, $h_l\not=0$ in which we are interested. These can be easily obtained by
inserting eq.(\ref{u_t}) into $\Phi_0 u_t^2 log(u_t)$ and then making the
suitable derivatives and limits~\cite{af}. In the case of the free energy one
obtains a term proportional to $h_l^4 log(h_l)$ which is too high to be
observed in our fits. However for the internal energy the first contribution is
proportional to a smaller power of $h_l$: $h_l^2 log(h_l)$ and must be taken
into account in the scaling function.

\end{description}

\subsection{The effective lattice Hamiltonian}

Let us call $H_{CFT}$ the Hamiltonian which describes the continuum theory at
the critical point. The perturbed Hamiltonian
\footnote{Here we follow the convention usually adopted in conformal field 
theory.
 In the standard notation of classical statistical mechanics one would  denote 
  this quantity 
 ``Hamiltonian density" rather than ``Hamiltonian".} 
in the continuum is given by:
\eq
H=H_{CFT}+h\sigma \;\;\;.
\en
The aim of this section is to construct the lattice analogous (which we shall
call $H_{lat}$) of $H$. 

Notice that $H_{lat}$ is different from 
the microscopic Hamiltonian which appears in the
exponent of eq.~(\ref{fupart}). Eq.~(\ref{fupart})
describes the model at the level of the lattice spacing. We are instead
interested in the large distance effective Hamiltonian
 which one obtains when the
short range degrees of freedom are integrated out, i.e. after a large enough
number of iterations of the Renormalization Group transformation has been
performed.  $H_{lat}$ will contain all the irrelevant operators which are
compatible with the symmetries of the lattice model. In this section we shall
first discuss the relation between the lattice and the continuum operators,
then we shall construct the lattice Hamiltonian in the $h_l=0$  case
and finally we shall extend our results to the $h_l\not= 0$ case.

\subsubsection{Relations  between lattice and continuum operators.}

The lattice operators are  given by the most general combination of
continuum operators compatible with the symmetries of the lattice operator
multiplied by the most general analytic functions of $t$ and $h_l$ 
(with a parity which
is again constrained by the symmetry of the operators involved). In the
following, to avoid a too heavy notation, we shall
neglect the $t$ dependence\footnote{The $t$ dependence in the scaling variables
of the irrelevant operators  plays a role only in the construction of the
scaling function for the internal energy and we shall resume it in sect.~3.4.3
below.}.

For the spin operator we have
\eq
\sigma_l=f^\sigma_0(h_l)\sigma+ h_l f^\epsilon_0(h_l)
\epsilon + f^\sigma_i(h_l) \sigma_i + h_l f^\epsilon_i(h_l)
\epsilon_i + h_l f^I_i(h_l)\eta_i,~~~  i\in{\bf N}
\label{xconvs}
\en
where $f^\sigma_i(h_l)$ $f^\epsilon_i(h_l)$ and $f^I_i(h_l)$ 
are even functions of $h_l$.

For the energy operator 
 we have
\eq
\epsilon_l=g^\epsilon_0(h_l)\epsilon + h_l g^\sigma_0(h_l)
\sigma + h_l g^\sigma_i(h_l) \sigma_i + g^\epsilon_i(h_l)
\epsilon_i +  h_l^2 g^I_i(h_l)\eta_i,~~~  i\in{\bf N}
\label{xconve}
\en
where again $g^\sigma_i(h_l)$ $g^\epsilon_i(h_l)$ and $g^I_i(h_l)$ 
are even functions of $h_l$ and the $h_l^2$ term in front of
$g^I_i(h_l)$ is due to the change of sign
 of the $\epsilon$ operator under duality
transformation at $h_l=0$ (see the discussion at the beginning of sect.~3.3.2).

 Among all the possible irrelevant fields
only those which respect the lattice symmetries (i.e. those of
 spin  $0~~(mod~4)$) are allowed in the sums. At this stage
 also irrelevant operators
containing $L_{-1}$ or $\bar L_{-1}$ appear in the sums. It is only when these
operators are applied on translationally invariant states 
(i.e. on the vacuum) that they disappear. This will 
 happen for instance when we shall study the  mean value of the free energy.

\subsubsection{ Construction of $H_{lat}(h_l=0)$}
In this case all the operators belonging to the $[\sigma]$ family are excluded
due to the $Z_2$ symmetry. Also the operators belonging to the $[\epsilon]$
family are excluded for a more subtle reason. The Ising model (both on the
lattice and in the continuum) is invariant under duality transformations while
the operators belonging to the $[\epsilon]$
family change sign under duality, thus they also cannot appear in
$H_{lat}(h_l=0)$. Thus we expect
\eq
H_{lat}~=~H_{CFT}~ +~ u^0_i~\eta_i, ~~~~~~~~~\eta_i \in [{I}]~~~,
\label{eq_rev}
\en
where the $u^0_i$ are constants.
There are however further restrictions:
\begin{itemize}
\item
 $H_{lat}$ is a scalar density, hence only operators $\phi$
 with angular momentum $j=4k$,  $k=0,1,2\cdots$ are allowed.
\item
 The operator which acts on the Hilbert space of the theory is the space
 integral of the Hamiltonian (density) $H_{lat}$. As such it is
 translational invariant and only
 operators $\phi$ which do not contain the generators $L_{-1}$ or 
$\bar L_{-1}$ survive the integration of eq.~(\ref{eq_rev}) over the space.
\end{itemize}
Let us list in order of increasing conformal weight the first few operators
which fulfill all the constraints:
\eq
\phi_0=I,~~~~
\phi_1=L_{-2}\bar L_{-2}I,~~~~
\phi_2=(L_{-2})^2I,~~~~
\phi_3=(\bar L_{-2})^2I,
\en
\eq
\phi_4=L_{-4}I,~~~~
\phi_5=\bar L_{-4}I,~~~~
\phi_6=L_{-3}\bar L_{-3}I~~~~... \;\;.
 \en
Some of these fields have a natural interpretation. 
$\phi_0$ gives rise to the bulk
contribution in the free energy (see the discussion of sect.~2.1.2).
$\phi_1,~\phi_2,~\phi_3$ are related to the energy momentum tensor: $T\bar T,
~T^2, {\bar T}^2$ respectively. All the fields listed above except the identity
and $\phi_6$ have the same conformal weight $h_\phi=4$.
 The corresponding RG eigenvalue is $y_\phi=-2$. The field $\phi_6$ has 
conformal weight $h_{\phi_6}=6$ and RG eigenvalue $y_{\phi_6}=-4$.

\subsubsection{Extension to $h_l\neq 0$}
Mimicking the continuum case we have, also on the lattice,
\eq
H_{lat}(h_l)=H_{lat}(h_l=0)+ h_l \sigma_l \;\;.
\en
Inserting the expression of $\sigma_l$ of eq.~(\ref{xconvs})
 we find
\eq
H_{lat}(h_l)~=~H_{CFT}~ +~ u_i(h_l)~\phi_i 
\en
where this time there is no more restriction coming from the $Z_2$ symmetry
and duality, hence $\phi_i$ denotes here the most general  operator
 of the spectrum with spin $j=4k$, $k=0,
1,..$
The $u_i(h_l)$ are even or odd functions of $h_l$, depending on the parity of
$\phi_i$ but in the even sector {\sl only} for the operators belonging to the
identity family $\lim_{h_l\to 0}u_i(h_l)\neq 0$ 
(according to eq.~(\ref{a3}) we have
$u_i(h_l=0)=u^0_i$ ).
For the operators belonging to the energy
family the first nonzero contribution in the $u_i(h_l)$ functions is of order
$h_l^2$.

Let us list, starting from those with the lowest conformal weight,
the new operators which were not present in $H_{lat}(h_l=0)$. For future
convenience let us separate those 
 which do not contain $L_{-1},\bar L_{-1}$ generators from the remaining 
ones.
\begin{description}
\item{A]}\hskip 0.5cm {\bf Operators which are not generated 
by $L_{-1},\bar L_{-1}$.}

\begin{itemize}
\item {\bf Operators belonging to $[\sigma]$}

In the $[{\sigma}]$ family the lowest ones are 
 $L_{-4}\sigma$,  $\bar L_{-4}\sigma$ and
 $L_{-3}\bar L_{-3}\sigma$.
In fact $L_{-1}\sigma$ disappears for translational invariance and due to the
null vector equation the $L_{-2}\sigma$
 operator which appears at level 2 can always be rewritten as $L_{-1}^2\sigma$
 with suitable coefficients.
The conformal weights of
$L_{-4}\sigma$ and $\bar L_{-4}\sigma$ are $h_{\sigma,4}=4+\frac18$.
 The corresponding 
RG eigenvalue is $y_{\sigma,4}=-2-\frac18$.
The conformal weight of
$L_{-3}\bar L_{-3}\sigma$ is $h_{\sigma,3\bar3}=6+\frac18$. The corresponding 
RG eigenvalue is $y_{\sigma,3\bar3}=-4-\frac18$.
\item
{\bf Operators belonging to $[\epsilon]$}

The most important contribution from the $[\epsilon]$ family is the one
proportional to $h_l^2\epsilon$ which is responsible for the $h_l^2$ term which
appears in $u_t$ as we discussed in the previous section.
Besides this one,
the lowest operators which appear in the $[{\epsilon}]$ family  must be
of the type $L_{-4}\epsilon$ or $\bar L_{-4}\epsilon$.
In fact the same mechanism which allowed us to eliminate the secondary fields of
 level 2 in the $[\sigma]$ family also works for the $[\epsilon]$ family.
On top of this
in the $[\epsilon]$ family a new null vector appears at level 3, thus
allowing us to eliminate also all the fields at this level. Keeping also into
account the fact that the corresponding $u_i(h_l)$ functions must start from
$h_l^2$ we immediately see that all these operators
 have too high powers of $h_l$ to
contribute to the scaling function and can be neglected.

\end{itemize}
\item{B]}\hskip 0.5cm 
{\bf Operators which contain $L_{-1},\bar L_{-1}$ generators.}

The lowest operators are, in order of increasing weight:
\begin{itemize} 
\item $L_{-1}\bar L_{-1}\sigma$, whose conformal weight is
 $h_{\sigma,1\bar1}=2+\frac18$. The corresponding 
RG eigenvalue is $y_{\sigma,1\bar1}=-\frac18$.

\item $L^2_{-1}\bar L^2_{-1}\sigma$, whose conformal weight is
 $h_{\sigma,2\bar2}=4+\frac18$. The corresponding 
RG eigenvalue is $y_{\sigma,2\bar2}=-2-\frac18$.

\item $L_{-1}\bar L_{-1}\epsilon$, whose conformal weight is
 $h_{\epsilon,1\bar1}=3$. The corresponding 
RG eigenvalue is $y_{\epsilon,1\bar1}=-1$.

\end{itemize}

\end{description}

\subsection{Scaling functions}
Using the results of the previous section we are now 
in the position to write the
 expression for the scaling functions keeping all the corrections up to the
 order $h_l^3$. 
\subsubsection{The free energy}
Due to translational invariance,
 only the secondary fields which are not generated by 
 $L_{-1},\bar L_{-1}$ contribute to the free energy.
We find, 
for the singular part of the lattice free energy:

\eqa
f_s(h_l)&=&~A^l_f~|h_l|^{\frac{16}{15}}(1+A^l_{f,1}|h_l|^{\frac{16}{15}}+
A^l_{f,2}|h_l|^{\frac{22}{15}}+
A^l_{f,3}|h_l|^{\frac{30}{15}}+
A^l_{f,4}|h_l|^{\frac{32}{15}}+ \nonumber \\
&&
A^l_{f,5}|h_l|^{\frac{38}{15}}+ 
A^l_{f,6}|h_l|^{\frac{44}{15}}
..... )
\label{f0}
\ena
where $A^l_{f,n}$ denotes the amplitude,
 normalized to the critical amplitude, of
the $n^{th}$ subleading correction.

\vskip 0.3cm
Let us discuss the origin of the various corrections:
\begin{itemize}

\item
${\bf A^l_{f,1}|h_l|^{\frac{16}{15}}}$

this term is entirely due to the $T\bar T$, $T^2$ and $\bar T^2$ irrelevant
fields in the Hamiltonian.
\item
${\bf A^l_{f,2}|h_l|^{\frac{22}{15}}}$

this term is  due to the $b_th^2_l$ term in $u_t$ (or, equivalently, to
the appearance of a $h^2\epsilon$ term in the Hamiltonian).

\item
${\bf A^l_{f,3}|h_l|^{\frac{30}{15}}}$

this term is  due to the $e_hh^2_l$ term in $u_h$.

\item
${\bf A^l_{f,4}|h_l|^{\frac{32}{15}}}$

this term keeps into account the second term in the Taylor expansion of the
$T\bar T$ like corrections and the contribution of the fields
$L_{-3}\bar L_{-3} I$ and $h L_{-4}\sigma$ in the Hamiltonian.

\item
${\bf A^l_{f,5}|h_l|^{\frac{38}{15}}}$

this is the product of the $A_{f,1}$ and $A_{f_2}$ corrections.

\item
${\bf A^l_{f,6}|h_l|^{\frac{44}{15}}}$

this is the second term in the Taylor expansion of the $h_l^2\epsilon$
correction.

\end{itemize}

To these terms we must then add the bulk contributions 

\eq
f_b(h_l)= f_b+f_{b,1}h_l^2+f_{b,2}h_l^4+\cdots \;\;\;.
\label{fx}
\en

We have already noticed in sect.~2.1.2 that $f_b$ can be obtained from the
exact solution of the Ising model on the lattice at the critical point. Also
the next term: $f_{b,1}$ can be evaluated (with a precision of
 ten digits) by noticing that it corresponds to the constant contribution to
 the susceptibility at the critical point. This term has been evaluated
 in~\cite{kap}. We  neglect for the moment this information and keep
the $f_{b,1}$ amplitude in the scaling function as a free parameter. It is the
first subleading term in the scaling function and as such it can be rather
precisely estimated with the fitting procedure that we shall discuss below. We
shall compare our estimates with the expected value in sect.~6 and use the
comparison as a test of the reliability of our results.

Combining eqs.~(\ref{f0}) and (\ref{fx})  we find:
\eqa
f(h_l)&=&~ f_b~+  A^l_f~|h_l|^{\frac{16}{15}}(1+
A^l_{f,b}|h_l|^{\frac{14}{15}}+
A^l_{f,1}|h_l|^{\frac{16}{15}}+
A^l_{f,2}|h_l|^{\frac{22}{15}}+  \nonumber \\
&&
A^l_{f,3}|h_l|^{\frac{30}{15}}+ 
A^l_{f,4}|h_l|^{\frac{32}{15}}+
A^l_{f,5}|h_l|^{\frac{38}{15}}+
A^l_{f,6}|h_l|^{\frac{44}{15}}
..... )
\label{f1bis}
\ena

where $A^l_{f,b}$ is $\frac{f_{b,1}}{A^l_f}$
and $A^l_{f,6}$ takes also into account now the contribution of $f_{b,2}$.

Deriving this expression with respect to $h_l$ we obtain the scaling functions
for the magnetization and the susceptibility\footnote{In this way 
we obtain directly the
lattice definitions of these two quantities, since we are deriving the lattice
free energy with respect to the lattice magnetic field. There is no need to
go  through the continuum definition of the magnetization.}.

\subsubsection{Mass spectrum}
The simplest way to deal with the mass spectrum is to 
 fit the {\sl square} of the masses. The scaling function turns out to be
 very similar to that which describes the singular part of the free energy
eq.~(\ref{f0}). The only additional terms are due to
 the secondary fields which contain $L_{-1}\bar L_{-1}$. It turns out that the
 corresponding scaling dimension exactly match those which already appear in
 eq.~(\ref{f0}). In fact 
\begin{itemize}
\item 
 $h_l L_{-1}\bar L_{-1}\sigma$ gives a contribution which scales with
  $|h_l|^{\frac{16}{15}}$ and its amplitude can be absorbed in $A^l_{f,1}$.
\item 
  $h_l L^2_{-1}\bar L^2_{-1}\sigma$ gives a contribution which scales with
  $|h_l|^{\frac{32}{15}}$ and its amplitude can be absorbed in $A^l_{f,4}$.
\item 
  $h_l^2 L_{-1}\bar L_{-1}\epsilon$ gives a contribution which scales with
  $|h_l|^{\frac{38}{15}}$ and its amplitude can be absorbed in $A^l_{f,5}$.
\end{itemize}

Thus the functional form of the scaling function for the masses is exactly the
same of eq.~(\ref{f0}).
\eqa
m^2_i(h_l)&=&(A^l_{m_i})^2~|h_l|^{\frac{16}{15}}(1+A^l_{m_i,1}|h_l|^{
\frac{16}{15}}+
A^l_{m_i,2}|h_l|^{\frac{22}{15}}+
A^l_{m_i,3}|h_l|^{\frac{30}{15}}+
A^l_{m_i,4}|h_l|^{\frac{32}{15}}+  \nonumber \\
&&
A^l_{m_i,5}|h_l|^{\frac{38}{15}}+
A^l_{m_i,6}|h_l|^{\frac{44}{15}}
..... )
\label{scalmass}
\ena

 However we shall see below that the presence of these new fields
 and in particular of
 $L_{-1}\bar L_{-1}\sigma$ has very important consequences.

\subsubsection{Internal energy}

 We may obtain the internal energy as a derivative with respect to $t$ 
 of the singular part of the free energy. However in
 doing this we must resume (as discussed above)
 the $t$ dependence in the scaling variables. This leads to some new terms in
 the scaling function with powers $\frac{8}{15}$ (due to the $c_ht$ term in
 $u_h$), $\frac{24}{15}$ and 
$\frac{40}{15}$ (due to the $t$ terms in scaling variables of the irrelevant
operators). It is nice to see that the presence of these additional 
contributions can be
understood in another, equivalent, way. Looking at eq.~(\ref{new1}) or
(\ref{xconve}) we see that the internal energy on the lattice contains a term
of type  $h_l\sigma$. The powers listed above are exactly those that we 
obtain keeping into account the additional $h_l\sigma$ term 
 in the scaling function. Keeping also into account the bulk
 contribution we end up with the following scaling function.
We have:
\eqa
E(h_l)~&=&~A^l_E~|h_l|^{\frac{8}{15}}(1+
A^l_{E,1}|h_l|^{\frac{8}{15}}+
A^l_{E,2}|h_l|^{\frac{16}{15}}+
A^l_{E,b}|h_l|^{\frac{22}{15}}+
A^l_{E,log}|h_l|^{\frac{22}{15}}log|h_l|+
\nonumber \\
&&
A^l_{E,3}|h_l|^{\frac{24}{15}}+
A^l_{E,4}|h_l|^{\frac{30}{15}}+ 
A^l_{E,5}|h_l|^{\frac{32}{15}}+
A^l_{E,6}|h_l|^{\frac{38}{15}}+
A^l_{E,7}|h_l|^{\frac{40}{15}}+
..... )
\label{eb}
\ena
where
 $A^l_{E,b}$ denotes the amplitude of the $h_l^2$ term in the bulk part of
the internal energy, 
 $A^l_{E,log}$ denotes the amplitude of the $h_l^2 log|h_l|$ term discussed in
 sect.3.1
and the bulk constant term has been already taken into
account in the definition of $E(h)$.
The first correction which appears in the internal
 energy (with amplitude $A^l_{E,1}$)  is
 the one with the lowest power of $h_l$
 among all the subleading terms of the
 various scaling functions this. Its effect on the scaling
 behaviour of the internal energy is very important and it is easily observable
 also in standard Monte Carlo simulations~\cite{cgm}. 

\subsubsection{Overlaps}

Also in this case we fitted the square of the overlap constants.
The scaling functions can be obtained with a straightforward application
 of the arguments discussed above. Also fields generated by
  $h L_{-1}\bar L_{-1}$ must be taken into account. Moreover, for the overlaps 
with the internal energy  also the $h\sigma$ term must be taken into account.
We end up with the following result for the magnetic
overlaps.

\eqa
|F_i^{\sigma}(h_l)|^2~&=&~|A^l_{F_i^\sigma}|^2(1+
A^l_{{F_i^\sigma},1}|h_l|^{\frac{14}{15}}+
A^l_{{F_i^\sigma},1}|h_l|^{\frac{16}{15}}+
A^l_{{F_i^\sigma},1}|h_l|^{\frac{22}{15}}+
\nonumber \\
&&
A^l_{{F_i^\sigma},1}|h_l|^{\frac{28}{15}}+
A^l_{{F_i^\sigma},1}|h_l|^{\frac{30}{15}}+
..... ) \;\;\;.
\label{fratio}
\ena

While for the  energy overlaps we have

\eqa
|F_i^{\epsilon}(h_l)|^2~&=&~|A^l_{F_i^\epsilon}|^2(1+
A^l_{{F_i^\epsilon},1}|h_l|^{\frac{8}{15}}+
A^l_{{F_i^\epsilon},1}|h_l|^{\frac{16}{15}}+
A^l_{{F_i^\epsilon},1}|h_l|^{\frac{22}{15}}+
\nonumber \\
&&
A^l_{{F_i^\epsilon},1}|h_l|^{\frac{24}{15}}+
A^l_{{F_i^\epsilon},1}|h_l|^{\frac{30}{15}}+
..... ) \;\;\;.
\label{fratio2}
\ena

\section{The transfer matrix method}
We computed the mass spectrum and observables by numerical diagonalization 
of the transfer matrix. The transfer matrix was introduced by Kramers and 
Wannier \cite{KrWa} in 1941. For a discussion of the transfer matrix  
see e.g. refs. \cite{CaFi,Ni90}.
The starting point is a simple transformation 
of the Boltzmann factor
\begin{equation}
\exp\left(\beta \sum_{<n,m>} \sigma_n \sigma_m + h_l \sum_n \sigma_n \right)
= T(u_1,u_2)\;T(u_2,u_3)  \; ... \; T(u_{L_0},u_1) 
\end{equation}
where $u_{n_0} = (\sigma_{(n_0,1)}, \sigma_{(n_0,2)}, \;...\; \sigma_{(n_0,L_1)})$
is the spin configuration on the time slice $n_0$.
$T$ is given by
\begin{equation}
 T(u_{n_0},u_{n_0+1})  =
 V(u_{n_0})^{1/2}  \;\;\;
			      U(u_{n_0},u_{n_0+1}) \;\;\;
			      V(u_{n_0+1})^{1/2}
\end{equation}
with 
\begin{equation}
U(u_{n_0},u_{n_0+1}) = \exp\left(\beta \; \sum_{n_1 = 1}^{L_1} \;
\sigma_{(n_0,n_1)} \sigma_{(n_0+1,n_1)} \right)
\end{equation}
and
\begin{equation}
V(u_{n_0}) = \exp\left(\beta \;\sum_{n_1 = 1}^{L_1} 
\sigma_{(n_0,n_1)} \sigma_{(n_0,n_1+1)} 
\;+\; h_l \;\sum_{n_1}^{L_1} \sigma_{(n_0,n_1)} \right) \;\;.
\end{equation}

The partition function becomes
\begin{equation}
Z \;=\; \sum_{\sigma_n \pm 1} 
\; \exp\left(\beta \sum_{<n,m>} \sigma_n \sigma_m + h_l \sum_n \sigma_n \right) 
\; = \; \mbox{tr} \; T^{L_0} \;= \; \sum_{i} \lambda_i^{L_0}
\end{equation}
where $T$ is interpreted as a matrix. The time-slice configurations are the 
indices of the matrix. The number of configurations on a time slice is
$2^{L_1}$. Therefore the transfer matrix is a $2^{L_1} \times 2^{L_1}$ matrix.
By construction the transfer matrix is positive and symmetric. The $\lambda_i$ 
are the eigenvalues of the transfer matrix.

\subsection{Computing observables}
Observables that are defined on a single time slice can be easily expressed
in the transfer matrix formalism. Let us discuss as examples
the magnetisation and the internal energy.
\begin{eqnarray}
 <\sigma_{1,1}> &=& \frac{\sum_{\sigma=\pm 1} 
 \exp\left(\beta \sum_{<n,m>} \sigma_n \sigma_m + h_l \sum_n \sigma_n \right) \;
  \sigma_{1,1} }{Z} \nonumber \\
  &=& \frac{ \mbox{tr} \; S \; T^{L_0} } { \mbox{tr} \;T^{L_0} }
  = \frac{ \sum_i \;  \lambda_i^{L_0} \; <i| S |i>} { \sum_i \lambda_i^{L_0} }
\label{Tobservable}
\end{eqnarray}
where $S$ is a diagonal matrix. The values on the diagonal are given 
by $\sigma_{1,1}$ on the configurations. ($S(u,u')=\delta(u,u') \; u^{(1)}$
where $u^{(1)}$ denotes $\sigma$ on the first site of the time slice).
The $|i>$ are normalized eigenvectors of $T$.

In the limit $L_0 \rightarrow \infty$ the expression simplifies to
\begin{equation}
 <\sigma_{1,1}> =  <0| S |0>
\end{equation}
where $|0>$ is the eigenvector with the largest eigenvalue.

The energy can be computed in a similar way. The diagonal matrix corresponding 
to the energy is given by 
\begin{equation}
 E(u,u') = \delta(u,u') \; u^{(1)} u^{(2)} \;\; .
\end{equation}
Note that we can only express the product of nearest neighbour spins in this
simple form if both spins belong to the same time slice.

In order to understand the relation of the mass spectrum with the 
eigenvalue  spectrum of the transfer matrix we have to compute correlation
functions with separation in time direction. 
The time-slice correlation function eq.~(\ref{slicecorr}) becomes in the 
limit $L_0 \rightarrow \infty$
\begin{equation}
<S_0 \; S_{\tau}> = \sum_i \exp(-m_i \; |\tau|) 
<0|\tilde S |i> \; <i|\tilde S |0> 
\end{equation}
with
\begin{equation}
 m_i = - \log \left(\frac{\lambda_i}{\lambda_0}\right)
\end{equation}
and $\tilde S= \frac{1}{L_1} \; \delta(u,u') \; \sum_{n_1} u^{n_1} $.
Note that $\tilde S$ is translational invariant (in the space direction)
 and has therefore only
overlaps with zero-moment eigenvectors of $T$. 

With eq.~(\ref{threes0}) we get
\begin{equation}
|F^{\sigma}_i| = \sqrt{m_i \; L_1} \; \frac{<0|\tilde S|i>}{<0|\tilde S|0>} \;\;\;.
\end{equation}

An analogous result can be obtained for the energy.

\subsection{Computing the eigenvectors and eigenvalues of $T$}
The remaining problem is to compute (numerically)  eigenvectors and eigenvalues
of the transfer matrix. Since we are interested in the thermodynamic limit as well
as in the continuum limit we would like to use as large values of
 $L_1$ as possible. This soon becomes a very difficult task
 since the dimension of the transfer matrix
 increases exponentially with $L_1$. The problem slightly simplifies if one is
 interested in the computation of the
the leading eigenvalues and eigenvectors only, and
in these last years various methods have been developed to address this
task (for a comprehensive discussion of existing approaches 
see e.g. ref. \cite{Ni90} or the appendix of ref. \cite{RiNoRi}).
In particular there are two approaches which have shown to be the most
effective ones.

\begin{itemize}
\item
 The first one reduces the numerical complexity of the problem 
by writing the 
transfer matrix as a product of sparse matrices. See refs. \cite{Ni90,RiNoRi}.
\item
 The second one is to reduce the dimension of the 
transfer matrix by restricting it to definite channels.
\end{itemize}

Since we are only interested in the zero-momentum states of the system 
we decided to follow the second approach and to
 compute the zero-momentum reduced transfer matrix. 
The zero-momentum reduced transfer matrix acts on the space of
equivalence classes of configurations on slices that 
transform into each other by translations. 

The matrix elements of the reduced transfer-matrix are given by
\begin{equation}
\tilde T(\tilde u, \tilde v) = 
(n(\tilde u) \;\; n(\tilde v))^{-1/2} \;\;
\sum_{u \in \tilde u} \sum_{v \in \tilde v}
                              T(u,v) \;\; 
			      = (n(\tilde u) / n(\tilde v))^{1/2}
  \sum_{v \in \tilde v} T(u,v) \;\;.
\end{equation}
where $n(\tilde u)$  is the number of configurations in $\tilde u$. 
For example for $L_0=20$ the dimension of the transfer matrix is reduced
from $1048576$ to $52488$.

Still the matrix is too large to save all elements of the matrix in the 
memory of the computer.  Therefore we applied an iterative solver and 
computed the elements of $\tilde T$ whenever they were needed.

As  solver we used a generalized power method as discussed in the 
appendix of ref. \cite{RiNoRi}.

The lattice sizes that we could reach in this way were large enough for
our purpose, thus we made no further effort to improve our method and 
 it is well possible that
our algorithm  might still not be the optimal one. 

We propose here, as a suggestion to the interested reader,
 some directions in which it could be improved.
\begin{itemize}
\item
One could try to mix the two strategies mentioned above and
 try to factorize the reduced transfer matrix as a product of sparse matrices.
However note 
that the complexity of the problem increases exponentially with the lattice
size. Therefore even a big improvement in the method would allow just to 
go up in the maximal $L_1$ by a few sites.
\item
One could study the transfer matrix 
along the diagonals of the square lattice.  Since the distance between two 
points on the diagonal is  $\sqrt{2}$, naively one could increase the 
accessible  lattice size by a factor of $\sqrt{2}$.
\end{itemize}

\section{Thermodynamic limit}

In order to take the thermodynamic limit we must know the finite size scaling
behaviour of the various observables  as a function of $L_1$. This is  a
very interesting subject in itself and several exact results have been obtained
in this context starting from the exact S-matrix solution and 
using  Thermodynamic Bethe Ansatz (TBA) techniques~\cite{Fateev,z91}.

For instance, it is possible 
to construct a large $L$ asymptotic
expansion for the finite size scaling (FSS)
of the energy levels  based only on
 the knowledge of the exact
S-matrix of the theory~\cite{h91,km}. 
Let us look to this FSS behaviour in more detail.

Let us define $\Delta m_a(L)$ as 
the deviation of the mass $m_a$ of the particle $a$ from its
asymptotic value:

\eq
\Delta m_a(L) \equiv m_a(L)-m_a(\infty) \;\;\;.
\en
Then in the large $L$ limit,  the shift 
(normalized to the lowest mass $m_1$) is dominated by an
 exponential decrease of the type

\eq
\frac{\Delta m_a(L)}{m_1} \sim -\frac{1}{8m_a^2} \sum_{b,
c}'\frac{\lambda^2_{abc}}{\mu_{abc}}\exp{(-\mu_{abc}L)}
\label{f1new}
\en
where
 the constants $\mu_{abc}$ and $\lambda_{abc}$ can be obtained
from the S-matrix and the
 prime in the sum of eq.~(\ref{f1new}) means that the sum must be
done only on those combinations of indices that fulfill the condition:
$|m^2_b-m^2_c|<m^2_a$. In particular the  $\mu_{abc}$ turn out to be of order
one, so that the FSS corrections are dominated (as one could naively expect)
 by a decreasing exponential of the type $\exp(-L_1/\xi)$ where the correlation
 length $\xi$ is the inverse of the lowest mass of the theory.

In principle we could use our data to test also the TBA predictions for the
FSS. However we
preferred to follow a different approach. We chose values of $h$ large enough
so as to fulfill the condition $L_1/\xi>>1$ for the largest values of $L_1$
that we could reach. In this way we could essentially neglect all the details
of the FSS functions and approximate them with a single exponential (or, 
in some cases,
 with a pair of exponentials). In order to study the FSS functions one
should choose smaller values of $h$. We plan to address this issue
in a forthcoming paper.
With our choice of $h$ we drastically simplify the FSS problem, however
nothing is obtained for free. 
The price we have to pay following this route is that we need to
know several terms in the scaling functions to fit such large values of $h$.
This explains the major effort that we devoted to this issue in sect.~3.

\subsection{Numerical extrapolation}
According to the above discussion,
for the extrapolation of our data to the thermodynamic limit we made no 
use of the quantitative theoretical results. We made only use of the 
qualitative result that the corrections due to the  finite $L_1$ vanish 
exponentially. 

We used as ansatz for the extrapolation either
\begin{equation}
 A(L_1) = A(\infty) \;+\; c_1 \; \exp(-L_1/z_1) 
\label{correction1}
\end{equation}
or 
\begin{equation}
 A(L_1) = A(\infty) \;+\; c_1 \; \exp(-L_1/z_1)  \;+\; c_2 \; \exp(-L_1/z_2) 
\label{correction2}
\end{equation}
where $A$ represents any of the quantities that we have studied.

In order to compute the free parameters $A(\infty)$, $c_1$ and $z_1$ or 
$A(\infty)$, $c_1$, $z_1$, $c_2$ and $z_2$ we solved numerically the 
system of equations that results from the lattice sizes 
$L_{1,max}$, $L_{1,max}-1$ and $L_{1,max}-2$ or 
$L_{1,max}$, $L_{1,max}-1$,  $L_{1,max}-2$, $L_{1,max}-3$ and $L_{1,max}-4$.

The error of $A(\infty)$ was estimated by comparing results where $L_{1,max}$ 
is the largest lattice size that is available and from $L_{1,max}'=L_{1,max}-1$.
Mostly ansatz (\ref{correction2}) was used to obtain the final result. 
In some cases however the numerical accuracy was not sufficient to resolve 
the second exponential term. Then the final result was taken from the ansatz 
(\ref{correction1}).

\begin{table}[h]
\vskip 0.2cm
\begin{tabular}{|r|l|l|l|}
\hline
 $ L_1$ &      $f$  & $f$, eq.~(\ref{correction1})& $f$, eq.~(\ref{correction2}) \\
\hline
  4 & 0.993343441146 &                  &                    \\
  5 & 0.992384038449 &                  &                    \\
  6 & 0.992160642059 &  0.992092835647  &                    \\
  7 & 0.992102865951 &  0.992082710940  &                    \\
  8 & 0.992086845804 &  0.992080699493  &  0.992080279141    \\
  9 & 0.992082188258 &  0.992080279123  &  0.992080180502    \\
 10 & 0.992080787928 &  0.992080185903  &  0.992080161487    \\
 11 & 0.992080356320 &  0.992080164020  &  0.992080157709    \\
 12 & 0.992080220728 &  0.992080158619  &  0.992080156931    \\
 13 & 0.992080177480 &  0.992080157225  &  0.992080156758    \\
 14 & 0.992080163514 &  0.992080156853  &  0.992080156721    \\
 15 & 0.992080158958 &  0.992080156752  &  0.992080156716    \\
 16 & 0.992080157458 &  0.992080156722  &  0.992080156709    \\
 17 & 0.992080156961 &  0.992080156715  &  0.992080156713    \\
 18 & 0.992080156795 &  0.992080156712  &  0.992080156710    \\
 19 & 0.992080156739 &  0.992080156710  &  0.992080156710    \\
 20 & 0.992080156721 &  0.992080156712  &  0.992080156713    \\
 21 & 0.992080156714 &  0.992080156710  &        -           \\
\hline
\end{tabular}
\vskip 0.2cm
\caption{\sl Extrapolation of the free energy at $h_l=0.075$
to the thermodynamic limit.
In the first column we give the lattice size $L_1$. In the second  column
the free
energy for this lattice size is given. In the third column we present the 
extrapolation with a single exponential and in the fourth column the 
extrapolation with a double exponential ansatz.}
\label{extra}
\end{table}

In tab. \ref{extra} we give, as example, the extrapolation to the 
thermodynamic limit of the free energy
at $h_l = 0.075$. As input for the extrapolation we used the free energy
computed up to 12 digits. We consider all these digits save of rounding errors.
Within the given precision the free energy has not yet converged at $L_1=21$.
The single exponential extrapolation~(\ref{correction1})
converges (within the given precision)
at $L_1=18$. For larger lattices the result fluctuates in the last digit 
due to rounding errors
of the input data. The double exponential extrapolation~(\ref{correction1})
converges at $L_1=16$. As final result for the thermodynamic limit we quote
$f(0.075)=0.99208015671$.

\section{Analysis of the results}
The major problem in extracting the continuum limit results from the data
listed in tabs.~\ref{tab10}-\ref{tab13}
 is to estimate the systematic errors involved in the
truncation of the scaling functions that we use in the fits. 
We shall devote the
first part of this section to a detailed description of the procedure that we
followed to estimate this uncertainty. We shall give upper and lower bounds for
the critical amplitudes which turn out to be very near to each other and allow
for high precision predictions (in some cases 
 we can fix 5 or even 6 significative digits). We then 
 compare our predictions with the results obtained in the
framework of the S-matrix approach. In all cases 
 we find a perfect agreement
within our bounds.
 Finally in sect.~6.3, we give, {\sl assuming}
 as fixed input the S-matrix predictions for the critical amplitudes,  our
 best estimates for the amplitudes of some of the subleading terms involved in
 the fits. 
\subsection{Systematic errors}
In order to estimate the systematic errors involved in our estimates of the
critical amplitudes we performed for each observable several independent fits
starting with a fitting function containing only the dominant scaling dimension
and then adding the subleading fields one by one. For each fitting function we
tried first to fit all the exiting data (those listed in 
tabs.~\ref{tab10}-\ref{tab13}) and then
eliminated the data one by one  starting from the farthest from the critical
point (i.e. from those with the highest values of $h_l$).
 Among the (very 
large) set of estimates of the critical amplitudes we selected only those
fulfilling the following requirements:
\begin{description}
\item{1]} \phantom{X}
The reduced $\chi^2$ of the fit must be of order unity
\footnote{This is a slightly
incorrect use of the $\chi^2$  function since the input data are
affected by errors which are of systematic more than statistic nature.
Notice however that we do not use it to determine best fit values for the
observables that we fit (we shall only give upper and lower bounds for them) 
but only as a tool to eliminate those situations in which the 
 fitting functions are clearly unable to describe the input data.}.
 In order to fix
precisely a threshold we required the fit to have a confidence level larger than
$30\%$.
\item{2]} \phantom{X}
The number of degrees of freedom of the fit (i.e. the number of data fitted
minus the number of free parameters in the fitting function) must be larger
than 3.
\item{3]} \phantom{X}
For all the subleading fields included in the fitting function, the amplitude
estimated from the fit must be larger than the corresponding 
errors, otherwise the field is eliminated from the fit.
\item{4]} \phantom{X}
The amplitudes of the subleading fields (in units of the critical 
amplitude) must be 
such that when multiplied for the corresponding power of $h_l$, 
(for the largest value of $h_l$ involved in the fit)  must give a contribution
 much smaller than 1 (in order to fix a threshold we required it to be strictly
smaller than 0.3). 
\end{description}
In general only a small number 
 of combinations of data and degrees of freedom
fulfills simultaneously all these requirements. Among all the corresponding
estimates of the critical amplitude we then select the smallest and the largest
ones as lower and upper bounds\footnote{In 
making this choice we also keep into account the errors in the
estimates induced by the systematic errors of the input data.}. 

\subsection{Critical amplitudes}
In tab.~\ref{tab7} we report as an example the fits to the magnetization 
(with the scaling
function obtained by deriving eq.~(\ref{f1bis})) fulfilling the
above requirements. For each value of $N_f$ we only report the fits with the 
minimum and maximum allowed number of d.o.f., since the best fit result for
 $A_M^l$ changes monotonically as the data are eliminated from the fit. This is
 a general pattern for all the observables that we studied and greatly 
simplifies the analysis of the data. Looking at the table one can see that at
least four parameters are needed in the fit to have a reasonable confidence
level, due to the very small error of the data that we use. In the last line 
we report the only fit in which
 all the 25 data reported in tab.~\ref{tab10}
 have been used. It required taking into
 account the first eight terms of the scaling function. 
For $N_f>8$, even if we use
 all the data at our disposal  we cannot fulfil requirement 3. 
It is interesting to notice that the fits which give the best approximations to
the exact value of $A_M^l$ are those in which we use the largest possible
number of terms of the scaling function. This is a general pattern for all the
observables that we studied.
All the fits were performed using the double precision NAG routine GO2DAF.
The bounds that we obtained are listed in tab.~\ref{tab8} together with
the S-matrix predictions. From these results we immediately obtain the upper
and lower bounds for the universal amplitude ratios of tab.~\ref{tab2}.
They are reported in the last two lines of tab.~\ref{tab8}.

\begin{table}[h]
\vskip 0.2cm
\begin{tabular}{|l|l|l|l|}
\hline
   $A_M^l$     &   $N_f$        &  d.o.f. &  C.L.     \\
\hline
$1.05898893(196)$    & 4  & 4  & $83\%$ \\
$1.05899447(58)$    & 4  & 6  & $50\%$ \\
\hline
$1.05898584(74)$    & 5  & 6  & $94\%$ \\
$1.05898882(22)$    & 5  & 7  & $48\%$ \\
\hline
$1.05898178(156)$    & 6  & 6  & $98\%$ \\
$1.05898375(8)$    & 6  & 10  & $98\%$ \\
\hline
$1.05898433(8)$    & 7  & 14  & $80\%$ \\
\hline
$1.05898694(18)$    & 8  & 15  & $99\%$ \\
$1.05898729(13)$    & 8  & 17  & $96\%$ \\
\hline
\end{tabular}
\vskip 0.2cm
\caption{\sl Fits to the magnetization fulfilling the requirements 1-4 
(see text). In the first column the best fit results for the critical amplitude
(with in parenthesis the  error induced by the systematic errors of the input
data),
in the second column the number of parameters in the fit, in the third column
the number of degrees of freedom and in the last column the confidence level.
For each value of $N_f$ we only report the fits with the minimum and maximum
allowed number of d.o.f, since the best fit of $A_f^l$ changes monotonically as
the data are eliminated from the fit.}
\label{tab7}
\end{table}

\begin{table}[h]
\vskip 0.2cm
\begin{tabular}{|l|l|l|l|}
\hline
   Observable     &   Lower bound         &   Upper bound &  Theory     \\
\hline
$A^l_f$    & 0.9927985  & 0.9928005  & 0.9927995... \\
$A^l_M$    & 1.058980  & 1.058995  & 1.058986... \\
$A^l_\chi$  & 0.07055   & 0.07072   & 0.070599... \\
$A^l_E$    & 0.58050  & 0.58059  & 0.58051... \\
$A^l_{m_1}$& 4.01031  & 4.01052      & 4.01040... \\
$A^l_{m_2}$& 6.486  & 6.491      & 6.4890... \\
$A^l_{m_3}$& 7.91  & 8.02      & 7.9769... \\
$|A^l_{F_1^\sigma}|$& 0.6405  & 0.6411      & 0.6409... \\
$|A^l_{F_1^\epsilon}|$& 3.699  & 3.714     & 3.7066... \\
$|A^l_{F_2^\sigma}|$& $0.3$  & $0.35$      & 0.3387... \\
$|A^l_{F_2^\epsilon}|$& $3.32$  & $\sim 3.45$     & 3.4222... \\
\hline
$R_\chi$& $6.7774$  & $6.7789$     & 6.77828... \\
$ Q_2$& $3.2296$  & $3.2374$     & 3.23514... \\
\hline
\end{tabular}
\vskip 0.2cm
\caption{\sl Lower  and upper bounds for various critical amplitudes discussed
in the text and, in the last two lines,
 for the two universal amplitude ratios $R_\chi$ and $Q_2$.} 
\label{tab8}
\end{table}

\subsection{Subleading operators}

In principle we could try to estimate in the fits discussed above also
the amplitudes of the first two or three subleading terms in the
scaling functions, however it is clear that the results that we would
obtain would be strongly
cross correlated and we would not be able to give reliable estimates 
for the corresponding errors (except, at most, for the first
one of them, the next to leading term in the scaling function).

In order to obtain some information on the subleading terms we decided to
follow another route.
The results of the previous section strongly support the correctness 
of the S-matrix predictions. We decided then to assume
 these predictions
 as an input of our analysis, fixing their values in the scaling functions.
Then we  used the same procedure discussed in sect.~6.1 to identify the
amplitude of the first subleading field.
Let us look to the various scaling functions in more detail
 
\subsubsection{ Free energy}

This is the case for which we have the most precise data. Moreover we may use
the data for the magnetization and the susceptibility as a cross check of our
estimates.

 Combining all the data at our disposal
 we end up with a rather precise estimate
for $A^l_{f,b}$, which turns out to be bounded by:

\eq
-0.055~ <~ A^l_{f,b}~ <~ -0.050 \;\;\;.
\en

As mentioned in sect.~3.4.1 it is possible to evaluate this amplitude in a
completely different way, by looking at the constant term in the magnetic
 susceptibility of the model at the critical point. The comparison between our
 estimate and the expected value represents a test of the reliability of our
 fitting procedure.
The expected value of this amplitude~\cite{kap} is (in our units) 
\eq
 A^l_{f,b}~ =~ -0.0524442... 
\label{ampli1}
\en
which is indeed in perfect agreement with our estimates.

We can then use the value of eq.~(\ref{ampli1})
as a fixed  input and try to estimate the amplitude
of the following subleading field which has a very important physical meaning
being the contribution due to the presence of the $T\bar T$ (and related terms)
operator in the lattice Hamiltonian. Remarkably enough, it turns out, by
applying the usual analysis, that the corresponding amplitude $A_{f,1}^l$ is
compatible with zero. 
 More precisely we see that, changing the number of input data and of
 parameters in the scaling
 function, the sign of $A_{f,1}^l$ changes randomly and its modulus is
 never larger than $10^{-4}$. The same pattern is reproduced in the
 magnetization and in the susceptibility. We summarize these observations 
with the following bound
\eq
|A_{f,1}^l|<0.00005~~~~.
\label{tt1}
\en
This result agrees with the observation concerning the absence
of corrections due to irrelevant operators in the case $t \ne 0$ and $h=0$. 
(For a  thorough discussion of this point see~\cite{ssv1} and refs. therein.)

If we also assume that $A_{f,1}^l=0$ then we may give a reliable
estimate for the amplitude of $A_{f,2}^l$
which turns out to be bounded by:

\eq
0.020~ <~ A^l_{f,2}~ <~ 0.022 \;\;\;.
\en

This is the highest subleading
 term that we could study with a reasonable degree of
confidence in our scaling functions.

\subsubsection{Internal energy}
In the case of the internal energy the first subleading amplitude can be studied
with very high confidence since it is associated to a very small exponent:
$|h_l|^\frac{8}{15}$. The result turns out to be
\eq
-0.646~ <~ A^l_{E,1}~ <~ -0.644 \;\;\;.
\en
In this case the fits are so constrained that we can study with a rather good
degree of confidence also the next subleading correction, $A_{E,2}^l$
 which is very interesting, since it 
again contains the $T\bar T$ term discussed above.
In agreement with the previous observations also in this case the amplitude
turns out to be compatible with zero. More precisely its sign changes randomly
as the input data are changed in the fits and its modulus can be bounded by:
\eq
|A_{E,2}^l|<0.005~~~~.
\en
which is not as strong as the bound of eq.~(\ref{tt1}) 
but clearly goes in the same direction.

\subsubsection{Masses}
The most interesting feature of the 
scaling functions for the masses is that
 there is no analytic term and the first subleading
contribution $A^l_{m_i,1}$ is the exact analogous of the  
$A^l_{f,1}$ term for the
free energy.

In this case
we find a non zero contribution for $A^l_{m_i,1}$. In particular we find the
following bounds for the three masses that we studied:
\eq
-0.21~<~A^l_{m_1,1}~<~-0.20 \nonumber
\en
\eq
-0.48~<~A^l_{m_2,1}~<~-0.41 \nonumber
\en
\eq
-0.65~<~A^l_{m_3,1}~<~-0.50 \;\;\; .\nonumber
\en

In the case of the masses a preferred direction is singled out. Therefore, 
one has to expect that there is a finite overlap with the irrelevant
operator that breaks the rotational symmetry.
Notice that a similar contribution has been
observed  also in the case of the thermal perturbation of the Ising
model in~\cite{cprv} where the authors studied
the breaking of rotational invariance in the two point correlator
(see sect.IV-G of~\cite{cprv} for a discussion of this point).

Our results on the amplitude of the subleading corrections are summarized in
tab.~\ref{tab9}.
\begin{table}[h]
\begin{tabular}{|lll|}\hline
$-0.055~$ & $<~A^l_{f,b}~$ &             $<~~-0.050  $ \\
  &$ \phantom{<~}|A_{f,1}^l|~$ &                     $<~~\phantom{-}0.00005 $ \\
$\phantom{-}0.020~$ & $<~A^l_{f,2}~ $ &             $<~~\phantom{-}0.022 $ \\
$-0.646~$ & $<~A^l_{E,1}~ $ &            $<~~-0.644 $ \\
 & $\phantom{<~}|A_{E,2}^l|~$ &                      $<~~\phantom{-}0.005 $ \\
$-0.21~$ & $<~A^l_{m_1,1}~$ &            $<~~-0.20 $ \\ 
$-0.48~$ & $<~A^l_{m_2,1}~$ &            $<~~-0.41  $ \\
$-0.65~$ & $<~A^l_{m_3,1}~$ &            $<~~-0.50  $ \\
 \hline
\end{tabular}
\vskip 0.2cm
\caption{\sl Lower and upper bounds for  the amplitudes of some of 
the subleading corrections. }
\label{tab9}
\end{table}

\section{Conclusions}

The major goal of this paper was to test
 the S-matrix description proposed by
 Zamolodchikov in~\cite{z89} for
 the 2d Ising model perturbed by a magnetic field. To this end we 
 developed some tools and obtained some results which are rather interesting
 in themselves. In particular
\begin{itemize}
\item
We improved the standard transfer matrix calculations by implementing
a zero momentum projection which allowed us to drastically
reduce the dimension of the matrix.
\item
We discussed in detail the relationship between continuum   and
lattice observables. 
\item
By using CFT results at the critical point we constructed the first 7-8 terms
of the scaling functions for various quantities on the lattice.
\end{itemize}

We could obtain in this way very precise numerical
estimates for several critical amplitudes (in some cases with 5 or even 
6 significative digits) and in all cases 
we found a perfect agreement between S-matrix
predictions and lattice results.

By assuming the S-matrix predictions as an input of our analysis we could
estimate some of the subleading amplitudes in the scaling functions. In one
case the value of the subleading amplitude was already known and again we found
a complete
agreement between theoretical prediction and numerical estimate.
 For the
remaining ones there is up to our knowledge no theoretical prediction.
They are collected in tab.~\ref{tab8} and represent the most interesting 
outcome of our analysis. We leave them as a challenge for theorists working in
the field. 

Among the others, the most surprising result concerns the $T\bar T$ term which
turns out to have a negligible amplitude in the scaling functions of the 
 translationally invariant observables. It would be nice to understand which
 is the reason of such behaviour.

Let us conclude by stressing that the techniques that we have developed can be
easily extended to the case in which a combinations of both thermal and 
magnetic perturbations is present. In this case the exact integrability is lost
and our numerical methods could help to test new approaches and suggest new
ideas.

\vskip 1cm
{\bf  Acknowledgements}
We thank A.B.~Zamolodchikov, Al.B.~Zamolodchikov, V.~Fateev, 
M.~Campostrini, A.~Pelissetto, P.~Rossi, E.~Vicari and R.~Tateo for useful
discussions and correspondence on the subject. In particular we are deeply
indebted with A.B.~Zamolodchikov for his help in the construction
of the scaling functions discussed in sect.~3.
This work was partially supported by the 
European Commission TMR programme ERBFMRX-CT96-0045.

\newpage
{\bf Tables of data}

\begin{table}[h]
\caption{\sl Data used in the fits}
\vskip 0.2cm
\begin{tabular}{|l|l|l|l|}
\hline
    $h_l$     &        $f$         &     $M$ &           $E$    \\
\hline
0.20              & 1.106272538601(1)  & 0.934113075978(1)  & 0.182495416253(1)
\\
0.19              & 1.096943627061(1)  & 0.931644255995(1)  & 0.179101587939(1)
\\
0.18              & 1.087640179593(1)  & 0.929017517063(1)  & 0.175544472125(1)
\\
0.17              & 1.078363862266(1)  & 0.926215008782(1)  & 0.171809915290(1)
\\
0.16              & 1.069116534998(1)  & 0.923215694344(1)  & 0.167881687799(1)
\\
0.15              & 1.059900287285(1)  & 0.919994540350(1)  & 0.163741028380(1)
\\
0.14              & 1.050717483321(1)  & 0.916521430645(1)  & 0.159366050850(1)
\\
0.13              & 1.041570819851(1)  & 0.91275968274(1)   & 0.154730958303(1)
\\
0.12              & 1.032463401585(1)  & 0.90866397795(1)   & 0.149804982192(1)
\\
0.11              & 1.023398841451(1)  & 0.90417740232(1)   & 0.14455091814(1)
\\
0.10              & 1.014381396853(1)  & 0.89922709483(1)   & 0.13892305302(1)
\\
0.09              & 1.00541615982(1)   & 0.89371763122(1)   & 0.13286414108(1)
\\
0.08              & 0.99650933082(1)   & 0.88752055778(1)   & 0.12630083230(1)
\\
0.075             & 0.99208015671(1)   & 0.88411094491(1)   & 0.1228010112(1)
\\
0.066103019026467 & 0.98424336850(1)   & 0.87741739906(1)   & 0.1161548337(1)
\\
0.055085849188723 & 0.97462849835(1)   & 0.86771621938(2)& 0.10703505648(2)
\\
0.05             &  0.97022834(1)     & 0.86255168(1)     & 0.10241966(1)
\\
0.044068679350978 & 0.96513182856(1)   & 0.8558157835(1) & 0.096641767(1)
\\
0.033051509513233 & 0.95578360408(2)& 0.840485633(1)  & 0.084469355(1)
\\
0.03             &  0.95322656(1)     & 0.83533709(5)  & 0.0806726(1) 
\\
0.022034339675489 & 0.9466343376(2) & 0.81901353(2)   & 0.0695436(1)
\\
0.02             &  0.94497330(2)  & 0.8139196(1)   & 0.0663409(2) 
\\
0.015            &  0.9409395(1)   & 0.7988985(1)   & 0.057595(1)
\\
0.01             &  0.936994(1)    & 0.77805(5)     & 0.047045(3)
\\
0.0088137358702   & 0.93607461(2)   & 0.771605(1)     & 0.044149(2)
\\
\hline
\end{tabular}
\label{tab10}
\end{table}

\begin{table}[h]
\caption{\sl Data used in the fits}
\vskip 0.2cm
\hskip -2cm
\begin{tabular}{|l|l|l|l|l|}
\hline
   $  h_l$            &   $1/m_1$         &   $1/m_2$   &  $1/m_3$        \\
\hline
0.20              & 0.59778522553(1)   & 0.37795775263(1)    & 0.310888(1) \\
0.19              & 0.61388448719(1)   & 0.38765653507(1)    & 0.318578(1) \\
0.18              & 0.63134670477(1)   & 0.39818995529(1)    & 0.326940(1) \\
0.17              & 0.65037325706(1)   & 0.40968266918(1) & 0.336077(2) \\
0.16              & 0.67120940172(1)   & 0.42228634593(5) & 0.346115(3) \\
0.15              & 0.69415734924(1)   & 0.43618773124(1) & 0.357209(3) \\
0.14              & 0.71959442645(1)   & 0.45161985381(4) & 0.369548(4) \\
0.13              & 0.74799884641(1)   & 0.4688779288(2)  & 0.38338(1) \\
0.12              & 0.77998715416(1)& 0.488342470(1)   & 0.3990(1) \\
0.11              & 0.81637015277(1)& 0.510513817(1)   & 0.4168(1) \\
0.10              & 0.85823913569(5)& 0.5360654(1)     & 0.4374(5) \\
0.09              & 0.9071039295(1) & 0.5659287(6)     & 0.4624(5) \\
0.08              & 0.965123997(1)  & 0.60144(1)       & 0.492(1) \\
0.075             & 0.998514180(1)  & 0.62189(1)       & 0.508(1) \\
0.066103019026467 & 1.067300500(2)  & 0.66405(5)       & 0.543(1) \\
0.055085849188723 & 1.17524158(3)   & 0.7305(1)        &      \\
0.05              & 1.237044(1)     & 0.768(1)         &    \\
0.044068679350978 & 1.322589(6)     & 0.82(1)          &   \\
0.033051509513233 & 1.54057(2)      &                 &  \\
0.03              & 1.6218(2)       &               &  \\
0.022034339675489 & 1.91(1)         &                &  \\
\hline
\end{tabular}
\label{tab11}
\end{table}

\begin{table}[h]
\caption{\sl Data used in the fits}
\vskip 0.2cm
\hskip -2cm
\begin{tabular}{|l|l|l|l|l|}
\hline
    $ h_l$            &     $|F_1^{\sigma}|^2$   &  $|F_2^{\sigma}|^2$ \\
\hline
0.20              & 0.29041938711(1) & 0.03800933(1) \\
0.19              & 0.29570405694(1) & 0.04039999(1) \\
0.18              & 0.30107729858(1) & 0.04291078(1)    \\   
0.17              & 0.30653975241(1) & 0.04554676(1)       \\
0.16              & 0.31209194307(1) & 0.04831337(1)      \\ 
0.15              & 0.31773424601(1) & 0.05121641(1)       \\
0.14              & 0.32346684419(1) & 0.05426214(1)       \\
0.13              & 0.3292896717(1)  & 0.05745711(3)       \\
0.12              & 0.3352023388(1)  & 0.0608082(1)        \\
0.11              & 0.3412040323(4)  & 0.0643227(2)        \\
0.10              & 0.3472933781(4)  & 0.068008(1)         \\
0.09              & 0.3534682486(5)  & 0.07187(1)          \\
0.08              & 0.359725487(1)   & 0.0759(1)           \\
0.075             & 0.362883627(1)   & 0.0780(2)           \\
0.066103019026467 & 0.368548928(2)   & 0.0818(5)           \\
0.055085849188723 & 0.3756378(4)     &                    \\
0.05              & 0.378934(5)      &                 \\
0.044068679350978 & 0.38280(2)       &                   \\
0.033051509513233 & 0.3899(1)        &                   \\
\hline
\end{tabular}
\label{tab12}
\end{table}

\begin{table}[h]
\caption{\sl Data used in the fits}
\vskip 0.2cm
\hskip -2cm
\begin{tabular}{|l|l|l|l|l|}
\hline
    $ h_l$            &     $|F_1^{\epsilon}|^2$   &  $|F_2^{\epsilon}|^2$ \\
\hline
0.20              & 11.7000114647(1)  & 8.0468067(5)  \\
0.19              & 11.8448924368(1)  & 8.3246112(5)  \\
0.18              & 11.9888157747(1)  & 8.6017424(5)  \\
0.17              & 12.1315853880(1)  & 8.8774943(5)     \\  
0.16             &  12.2729879270(1)  & 9.1511600(5)       \\
0.15               &12.4127893980(1)  & 9.422023(1)        \\
0.14              & 12.5507307560(1)  & 9.689348(2)        \\
0.13              & 12.6865220830(5)  & 9.952360(5)        \\
0.12              & 12.819834783(1)  & 10.21022(1)         \\
0.11              & 12.950290902(2)  & 10.46202(3)         \\
0.10              & 13.077448185(2)  & 10.7067(5)          \\
0.09              & 13.200778543(2)  & 10.943(3)           \\
0.08              & 13.31963596(3)   & 11.17(1)            \\
0.075            &  13.3771415(1)    & 11.28(1)            \\
0.066103019026467&  13.475815(5)     & 11.46(2)             \\
0.055085849188723&  13.59037(3)      & 11.6(5)             \\
0.05             & 13.6398(5)        &                   \\
0.044068679350978&  13.695(1)        &                   \\
0.033051509513233&  13.78(1)         &                   \\
\hline
\end{tabular}
\label{tab13}
\end{table}


\newpage

\begin{thebibliography}{99}

\bibitem{bpz} A.A. Belavin, A.M. Polyakov and A.B. Zamolodchikov,
Nucl. Phys. {\bf B241} (1984) 333.

\bibitem{z89}
A.B. Zamolodchikov, in "Advanced Studies in Pure Mathematics"
{\bf 19} (1989) 641; Int. J. Mod. Phys. {\bf A3} (1988) 743.

\bibitem{int}
A.B. Zamolodchikov, Al B. Zamolodchikov, Ann. Phys. {\bf 120} (1979) 253.

\bibitem{rev} G. Mussardo, Phys. Rep. {\bf 218} (1992) 215. 

\bibitem{hs} M. Henkel and H.Saleur, \JP{A22} (1989) L513.


\bibitem{bnw} V. Bazhanov, B. Nienhuis and S.O. Warnaar, \PL{B322}
(1994) 198.

 U. Grimm, B. Nienhuis \PR{E55} (1997) 5011.
\bibitem{bs97} M.T. Batchelor and K.A. Seaton, \JP{A30} (1997) L479.

 M.T. Batchelor and K.A. Seaton, \NP{B520} (1999) 697.

\bibitem{lr} P.G. Lauwers and 
V. Rittenberg, Phys. Lett. {\bf B233} (1989) 197,
and  preprint Bonn University BONN-HE-89-11 (unpublished).

\bibitem{destri} C. Destri, F. Di Renzo, 
E. Onofri, P. Rossi and G.P. Tecchiolli,
Phys. Lett. {\bf B278} (1992) 311.

\bibitem{dm} G. Delfino, G. Mussardo, { Nucl. Phys.} {\bf B455}
(1995) 724.

\bibitem{chp} M. Caselle, M. Hasenbusch and P. Provero, \NP{B556} (1999) 575.

\bibitem{ff} A.E. Ferdinand and M.E. Fisher, 
{ Phys. Rev.} {\bf 185} (1969) 832. 

\bibitem{cgm} M. Caselle, P. Grinza and N. Magnoli, hep-th/9909065.

\bibitem{cardy} J. Cardy, {\sl Scaling and 
Renormalization in Statistical Physics},
 Cambridge University Press 1996.


\bibitem{ahp} V. Privman, P.C. Hohenberg, A. Aharony, {\em Universal
Critical-Point Amplitude Relations}, in ``Phase transition and critical
phenomena'' vol.\,14, C. Domb and J.L. Lebowitz eds. (Academic Press 1991).

\bibitem{Fateev} V. Fateev, \PL{B324} (1994) 45.


\bibitem{flzz} V. Fateev, S. Lukyanov, A. Zamolodchikov and Al. Zamolodchikov,
\NP{B516} (1998) 652.

\bibitem{d98} G. Delfino, \PL{B419} (1998) 291.

\bibitem{ds} G. Delfino, P. Simonetti, { Phys. Lett.} {\bf B 383} (1996) 
450.

\bibitem{smilga} A.V. Smilga, \PR{D55} (1997) 443.


\bibitem{mccoy} B.M. McCoy and T.T. Wu, {\sl The two dimensional Ising Model},
(Harvard Univ. Press, Cambridge,1973).

B.M. McCoy, in {\sl Statistical Mechanics and Field Theory}, eds. V.V. Bazhanov
and C.J. Burden, World Scientific, 1995.



\bibitem{wu}
T.T. Wu, Phys. Rev. {\bf 149} (1966) 380.



\bibitem{h67} R. Hecht, \PR{158} (1967) 557.

\bibitem{dsz} P. di Francesco, H. Saleur and J-B. Zuber, \NP{B290} (1987) 527

\bibitem{af} A. Aharony and M.E.Fisher, \PR{B27} (1983) 4394

\bibitem{kap} X.P. Kong, H. Au-Yang and J.H.H. Perk
\PL{A116} (1986) 54.


\bibitem{KrWa}
H.A. Kramers and G.H. Wannier, Phys. Rev. {\bf 60} (1941) 252, ibid. p 263.

\bibitem{CaFi} W. J. Camp and M. E. Fisher, 
{ Phys. Rev.} {\bf B6} (1972) 946.

\bibitem{Ni90}
M.P. Nightingale, in  {\sl Finite Size Scaling and 
Numerical Simulation of Statistical Systems},
ed. V. Privman, World Scientific 1990.

\bibitem{RiNoRi}
 H. L. Richards, M. A. Novotny and P. A. Rikvold,
{ Phys. Rev.} {\bf B48} (1993) 14584.

\bibitem{z91}
  Al.B. Zamolodchikov, 
Nucl. Phys. {\bf B342} (1990) 695, and
 \PL{B253} (1991) 391.
\bibitem{h91}
M. Henkel, \JP{A24} (1991) L133.
\bibitem{km}
T.R. Klassen and E. Melzer,
\NP{B362} (1991) 329.

\bibitem{ssv1} J. Salas and A.D. Sokal, cond-mat/9904038v1.

\bibitem{cprv} M. Campostrini, A. Pelissetto, P. Rossi and E. Vicari,
\PR{E57} (1998) 184.






\end{thebibliography}
\end{document}